\documentclass[aps, prx, reprint, longbibliography, nofootinbib,superscriptaddress, floatfix]{revtex4-1}

\usepackage{slashed}
\usepackage{verbatim}
\usepackage[T1]{fontenc}
\usepackage{mathbbol}
\usepackage[dvipsnames]{xcolor}
\usepackage{orcidlink}

\usepackage[normalem]{ulem}
\usepackage[english]{babel}
\usepackage{lipsum}
\usepackage{physics}
\usepackage{dcolumn}
\usepackage{tensor}
\usepackage{comment}
\usepackage{graphicx,color,overpic,mathtools}
\usepackage{amsthm,amsmath,amssymb,mathrsfs}
\usepackage{braket,bm,bbm,setspace}
\usepackage{cancel}
\usepackage{float}
\usepackage{xargs}
\definecolor{myred}{RGB}{179, 27, 27}
\usepackage{hyperref}
\hypersetup{
    colorlinks=true,
    linkcolor=myred, 
    citecolor=myred, 
    urlcolor=myred  
 }

\newcommand{\E}{\mathcal{E}}
\newcommand{\Pres}{\mathcal{P}}
\newcommand{\Q}{\mathcal{Q}}
\newcommand{\N}{\mathcal{N}}

\begin{document}

\title{Superradiant amplification by rotating viscous compact objects}

\author{Jaime Redondo-Yuste\orcidlink{0000-0003-3697-0319}}
\email[]{jaime.redondo.yuste@nbi.ku.dk}
\affiliation{Center of Gravity, Niels Bohr Institute, Blegdamsvej 17, 2100 Copenhagen, Denmark}

\author{Vitor Cardoso\orcidlink{0000-0003-0553-0433}}
\affiliation{Center of Gravity, Niels Bohr Institute, Blegdamsvej 17, 2100 Copenhagen, Denmark}

\begin{abstract}
We study fluctuations of rotating viscous stars, using the causal relativistic hydrodynamics of Bemfica, Disconzi, Kovtun, and Noronha. We derive, in a slow-rotation approximation, a coupled system of equations describing the propagation of axial gravitational waves through the star, which couple to internal viscous modes. We show that rotating viscous stars amplify incoming low-frequency gravitational waves, a phenomenon which we argue to be universal. Superradiant amplification does not seem to trigger an instability for uniformly rotating stars, even if the object is compact enough to have light rings.
\end{abstract}

\maketitle

\noindent \textbf{\emph{Introduction.}} 
Compact, self-gravitating stars are not perfect fluids. In neutron stars, for example, dissipative effects are significant -- shear viscosity in cold stars, bulk viscosity in hot ones. These influence their dynamics~\cite{Redondo-Yuste:2024vdb, Boyanov:2024jge}, and provide clues about the dense matter in their interiors~\cite{Shibata:2017xht, Alford:2018lhf, Alford:2020lla, Most:2021zvc, Most:2022yhe, Chabanov:2023abq, Chabanov:2023blf, Chabanov:2024yqv, Pandya:2022pif, Pandya:2022sff,Ripley:2023lsq, Ripley:2023qxo, HegadeKR:2024agt, HegadeKR:2024slr}. Viscosity also affects their dynamical tidal deformability~\cite{Ripley:2023lsq, Ripley:2023qxo, HegadeKR:2024agt, HegadeKR:2024slr}, radial stability~\cite{Caballero:2025omv}, and their relaxation to equilibrium~\cite{Shibata:2017xht, Alford:2018lhf, Alford:2020lla}, leaving imprints on the gravitational wave emission from merging neutron stars. 

Additionally, there may be self-gravitating compact bodies composed wholly or partly of dark matter, or exotic stars that mimic black hole properties~\cite{Cardoso:2019rvt,Buoninfante:2024oxl,Bambi:2025wjx}. When rotating, these bodies can develop ergo-region instabilities~\cite{Friedman:1978ygc,Moschidis:2016zjy,Barausse:2018vdb,Brito:2015oca}, which dissipation may suppress. Rotating compact objects also experience an r-mode instability---a relativistic analogue of Rossby waves, unstable against the emission of gravitational waves~\cite{Andersson:1997xt, Lindblom:1998wf,Andersson:2000mf}. In neutron stars, viscosity may quench this instability~\cite{Andersson:1997xt,Lindblom:1999yk,Lindblom:2000gu,Owen:1998xg,Lindblom:2001hd,Bildsten:1999zn,Kraav:2024cus}, though a first principles analysis remains open (see preliminary work in Ref.~\cite{Pons:2005gb}). Whether viscosity stabilizes other rotating compact objects, or rather, renders them more unstable is, in essence, unknown. 

Further insight comes from wave propagation across viscous interfaces. Consider a sound wave crossing an interface between a moving viscous fluid and one at rest. An extension of Refs.~\cite{1957ASAJ...29..435R, 1954ASAJ...26.1015M}, shows that viscosity can amplify waves~\cite{Boyanov:2024jge}. The mechanism behind this is the same as that triggering superradiant amplification of gravitational waves around rotating black holes~\cite{zeldovich1,zeldovich2,Brito:2015oca}. This suggests that rotating, viscous compact bodies may generically amplify low-frequency radiation. Such amplification competes with ergoregion and r-mode instabilities, shaping the stability of rotating stars and black hole mimickers.

In this Letter, we show that compact rotating stars amplify low-frequency radiation whenever viscosity is present. We derive a coupled system of equations describing the propagation of axial-driven modes in slowly rotating stars, using the causal and stable hydrodynamics of Bemfica, Disconzi, Noronha, and Kovtun (BDNK)~\cite{Bemfica:2017wps, Bemfica:2019knx, Bemfica:2020zjp, Kovtun:2019hdm, Hoult:2020eho}, which we integrate numerically to find superradiant amplification. Finally, by analyzing an extension of Zel`dovich's toy model, we argue that this amplification does not trigger a superradiant instability, since the wavelengths subject to amplification are not trapped efficiently, even for extremely compact objects. 

\noindent \textbf{\emph{Superradiant amplification.}} 
Objects with free microstates (internal degrees of freedom) capable of absorbing radiation, will amplify low-frequency waves at the cost of their rotational energy~\cite{zeldovich2,Bekenstein:1973mi,Brito:2015oca}. The thermodynamic argument considers any {\it axi-symmetric} macroscopic body rotating rigidly with constant angular velocity about its symmetry axis, and with well-defined entropy $S$, rest mass $M$, and temperature $T$. Suppose now that a wave packet with frequency $(\omega,\omega+d\omega)$ and azimuthal number $m$ is incident upon this body, with a power $P_m(\omega)d\omega$. Radiation with a specific frequency and azimuthal number carries angular momentum at a rate $(m/\omega)P_m(\omega)d\omega$ (see Appendix C in Ref.~\cite{Brito:2015oca}). Neglecting spontaneous emission by the body (of thermal or any other origin), it will absorb a fraction $Z_m$ of the incident energy and angular momentum (where the dot stands for time derivative),
\begin{equation}
\dot{E}= Z_m P_md\omega ,\quad  \dot{J}= Z_m \frac{m}{\omega} P_m d\omega \, .
\end{equation}
Note that the assumption of axi-symmetry and stationarity implies that no precession or Doppler shifts occur during the interaction. Both the frequency and
multipolarity of the incident and scattered wave are the same. Now, in the frame co-rotating with the body, the change in energy is simply
\begin{equation}
dE_0 = dE-\Omega dJ = dE\Bigl(1-\frac{m\Omega}{\omega}\Bigr)\,,
\end{equation}
and thus the absorption process is followed by an increase in entropy, $dS = dE_0/T$, of
\begin{equation}
\dot{S} = \frac{\omega-m\Omega}{\omega T} Z_m P_m(\omega)d\omega\,.\label{eq:entropy}
\end{equation}
The second law of thermodynamics demands that $\dot{S} > 0$. Thus, superradiant amplification, corresponding to $Z_m < 0$, occurs whenever $\omega - m\Omega < 0$ for positive frequency modes $\omega > 0$. 

This discussion and our results question ad hoc attempts to give rotating black hole mimickers an ``absorption coefficient'' at low frequencies (e.g. Refs~\cite{Maggio:2017ivp,Maggio:2021uge} and many others). Frame-dragging effects, e.g., defining the reflectivity with respect to the frequency as seen from the point of view of the local, rotating frame $\omega \to \omega-m\Omega$, are necessary in order to avoid violating the second law of thermodynamics.
Any black hole mimicker that does not describe a fundamental microstate -- such as fuzzballs~\cite{Mathur:2005zp} or topological stars~\cite{Bah:2020ogh} -- should, naively, have a large entropy, on a similar scale as the black hole entropy, $S \sim K S_{\rm BH}$, with $K\lesssim 1$ a constant. According to the Kovtun-Son-Starinets bound~\cite{Kovtun:2003wp,Kovtun:2004de}, this implies that such an object must also have a large shear viscosity, at least $\eta \gtrsim K S_{\rm BH}/(4\pi R_S^2)$, with $R_S$ the radius of the object. Notably, if the constant $K\sim 1$, meaning the mimicker has an entropy comparable to that of a black hole, this lower bound is much higher than the viscosity found in realistic neutron stars. Thus, although shear viscosity may be negligible for most neutron stars, it should not be neglected nor overlooked when studying black hole mimickers.

\noindent\textbf{\emph{Setup.}} 
We consider the metric of a slowly rotating body~\cite{Hartle:1967he, Hartle:1968si}, sourced by a viscous fluid,
\begin{equation}\label{bg_metric}
    ds^2 =-e^{\nu(r)} dt^2+e^{\lambda(r)}dr^2+r^2d\Omega^2-2r^2\varpi(r)\sin^2\theta dtd\phi \, ,
\end{equation}
where $d\Omega^2$ is the area element of the round unit $2$-sphere. The mass aspect $M(r)$ is defined through $e^{-\lambda} = 1-2M/r$. The stress-energy tensor includes first-order gradients of the thermodynamic variables, and is given by~\cite{Bemfica:2020zjp}
\begin{equation}
    T_{ab} = \mathcal{E} u_a u_b + \mathcal{P}(g_{ab}+u_au_b) + \mathcal{Q}_a u_b+\mathcal{Q}_b u_a + \mathcal{T}_{ab} \, , 
\end{equation}
where $u_a$ is the fluid $4$-velocity, and
\begin{equation}\label{constitutive_relations}
    \begin{aligned}
        \E =& \rho + \tau_\E \Bigl[u^a\nabla_a\rho+(\rho+P)\nabla_au^a \Bigr] \, , \\
        \Pres =& p - \zeta \nabla^au_a + \tau_{\Pres}\Bigl[u^a\nabla_a\rho+(\rho+P)\nabla_au^a \Bigr] \, , \\
        \Q_a=& \tau_\Q(\rho+p)u^b\nabla_bu_a+\beta_\E \Pi_{ab}\nabla^b \rho + \beta_\N \Pi_{ab}\nabla^b n \, , \\
        \mathcal{T}_{ab} =& -2\eta \sigma_{ab} \, , 
    \end{aligned}
\end{equation}
where $\rho,p,n$ are the fluid rest energy density, pressure, and particle number density, and $\tau_{\E,\Pres,\Q},\eta,\zeta,\beta_{\E,\N}$ are transport coefficients. We will restrict to barotropic matter $p=p(\rho)$, with constant temperature and chemical potential. Then, the sound speed is $c_s^2 = dp/d\rho$, and $\beta_\E = c_s^2\tau_\Q$, with $\beta_\N=0$. Equilibrium solutions are characterized by a uniformly rotating fluid velocity profile 
\begin{equation}\label{bg_fluid_velocity}
    u = e^{-\nu/2}\Bigl(\partial_t+\Omega \partial_\phi\Bigr) \, ,
\end{equation}
where $\Omega$ is the (constant) angular velocity of the star. We will only consider linear terms in $\Omega/\Omega_K$, where $\Omega_K = M_S^{1/2}R_S^{-3/2}$ is the (Keplerian) mass-shedding frequency of the star. Up to that order, we recover from Einstein equations the usual equations for relativistic hydrostatic equilibrium
\begin{equation}\label{tov_equations}
    M' = 4\pi r^2 p  \, , \quad \nu'= -\frac{2e^\lambda}{r^2}\Bigl(M+4\pi r^3p\Bigr) \, , \quad 
    p' = \frac{\rho+p}{2}\nu' \, ,
\end{equation}
along with the frame-dragging equation, 
\begin{equation}\label{frame_dragging_eq}
    \varpi'' +4\Bigl[\frac{1}{r}-\pi r e^\lambda (\rho+p) \Bigr]\varpi'+16\pi e^\lambda(\rho+p)(\Omega-\varpi) = 0 \, . 
\end{equation}
Given central values for the density and an equation of state, Eqs.~\eqref{tov_equations} can be integrated outwards to find the star structure, its radius $R_S$ and its mass $M_S=M(R_S)$. Imposing $\varpi'(0)=0$ we can then integrate Eq.~\eqref{frame_dragging_eq} outwards. The exterior matches onto a slowly rotating spacetime with mass $M_S$, and $\varpi = 2J/r^3$, where the angular momentum of the star can be computed as $6J=-R_S^4 \varpi'(R_S)$. The slowly rotating approximation requires that this angular velocity is much smaller than the mass-shedding frequency, $\Omega \ll \Omega_K$.
We consider a polytropic equation of state of the form $p=\kappa \rho^{1+1/n}$, and will focus on a stellar model with  $\kappa = 700 \mathrm{km}^{2.5}$, $n=0.8$, and central density $\rho_c = 3 \times 10^{15}\mathrm{g}/\mathrm{cm}^{3}$. In that case, the total mass and radius of the star are, respectively, $M_S = 1.6M_\odot$ and $R_S=8.2\mathrm{km}$, and compactness $M_S/R_S  =0.288$.

We further follow Ref.~\cite{Boyanov:2024jge} to parametrize the transport coefficients, which ensures the Israel junction conditions at the surface of the star are satisfied to linear order in the perturbations. In particular, we choose $\eta = p R_S \hat{\eta}$, and $\tau_\Q = \rho^{-1} p R_S\hat{\tau}$, where the rest of the transport coefficients, as well as the BDNK constraints, are discussed in detail in Ref.~\cite{Boyanov:2024jge}. The system is causal as long as $\hat{\eta} \leq 3/4$ and $\hat{\tau} > 1+ C\hat{\eta}$, with $C$ a constant which depends on the equation of state. 

\noindent\textbf{\emph{Linearised perturbations.}}
Linear perturbations of slowly rotating stars were first studied in Refs.~\cite{Chandrasekhar:1991rmo,Kojima:1992ie}. In the presence of rotation, the $\ell$-th axial multipole couples to the $\ell \pm 1$-th polar multipole, and vice versa. However, to linear order in the rotation rate, we can focus on axial--led modes, neglecting this coupling~\cite{Pani:2013pma}, which will be, nevertheless, important beyond linear order. We consider a perturbation of the form 
\begin{equation}
    g_{ab} = \bar{g}_{ab} + h_{ab} \, , \quad u^a  = \bar{u}^a + \delta u^a \, , 
\end{equation}
where $\bar{g}_{ab}$ is the background metric, corresponding to Eq.~\eqref{bg_metric}, and $\bar{u}^a$ the background fluid velocity~\eqref{bg_fluid_velocity}. In the Regge-Wheeler gauge, axial metric perturbations are given by 
\begin{equation}
    h_{iA} = h^{\ell m}_i(r) X^{\ell m}_A(\theta,\phi) e^{-i\omega t} \, , \qquad i=0,1 \, ,
\end{equation}
where $A=2,3$ are the angular components, and the rest of the metric perturbation components vanish. Vector spherical harmonics are defined as $X^{\ell m}_A=\epsilon_{A}^BD_B Y_{\ell m}$, in terms of the covariant derivative of the round unit sphere, $D_A$. The fluid velocity perturbation can be expanded similarly
\begin{equation}
    \delta u^a = e^{-\nu/2}e^{-i\omega t}\Bigl(U X_\phi,0,\frac{Z}{r^2}X_\theta,\frac{Z}{r^2\sin^2\theta}X_\phi\Bigr) \, ,
\end{equation}
where $U = e^{-\nu}\Bigl[Z(\Omega-\varpi)+\Omega h_0\Bigr]$, so that $\delta(u_au^a) = 0$. 

Following the separation procedure outlined in~\cite{Kojima:1992ie, Pani:2013pma} (see also Supplemental Material), we eliminate $h_0$ in terms of $h_1$ and $Z$, and use the remaining equations to write two coupled wave equations. These describe the propagation of gravitational waves and fluid viscous modes. They are more simply written in terms of a gravitational master variable $\psi$, defined as 
\begin{equation}
    h_1 = re^{(\lambda-\nu)/2}\Bigl(1-\frac{m\varpi}{\omega}\Bigr) \psi \, ,
\end{equation}
and the tortoise coordinate is $dr_\star = e^{(\lambda-\nu)/2}dr$. The master equations can be written compactly as 
\begin{equation}\label{wave_eqs}
    \begin{aligned}
        \frac{d^2\psi}{dr_\star^2} +\Bigl(\omega^2c^{-2}_{\psi}-\mathcal{V}_{\psi}\Bigr)\psi =& C_{11} \frac{d\psi}{dr_\star} + C_{12}\frac{dZ}{dr_\star} + C_{13}Z \, , \\
        \frac{d^2Z}{dr_\star^2} +\Bigl(\omega^2c^{-2}_{Z}-\mathcal{V}_{Z}\Bigr)Z =& C_{21} \frac{dZ}{dr_\star} + C_{22}\frac{d\psi}{dr_\star} + C_{23}\psi \, ,
    \end{aligned}
\end{equation}
where the coefficients $C_{ij}$ are given in an accompanying \texttt{Mathematica} notebook. The propagation speed 
of each of the modes receives a purely imaginary correction with respect to their non-rotating value, $c^2_\psi = 1 + i\Delta$, and $c^2_Z = \eta[\tau_\Q (p+\rho)]^{-1}(1-i\Delta)$, with 
\begin{equation}
    \Delta = \frac{16\pi m r^2 e^{-\nu/2}}{\ell(\ell+1)}\Bigl(\Omega-\varpi\Bigr)\Bigl[(\rho+p)\tau_\Q-\eta\Bigr] \, .
\end{equation}
The non-rotating limit of Eqs.~\eqref{wave_eqs} recovers the equations of Ref.~\cite{Redondo-Yuste:2024vdb}, whereas the perfect fluid limit recovers the master equation of Ref.~\cite{Kojima:1992ie}. 

\noindent\textbf{\emph{Scattering of waves.}} 
We consider the scattering of gravitational waves of frequency $\omega$ off the star. The propagation in the exterior is described by the slowly rotating Regge-Wheeler equation 
\begin{equation}
    \frac{d^2\psi}{dr_\star^2} +\Bigl(\omega^2-\frac{4m\omega J_S }{r^3}-\mathcal{V}_{\rm RW}\Bigr)\psi = 0 \, , \
\end{equation}
where
\begin{equation}
    \begin{aligned}
        \mathcal{V}_{\rm RW} =& f(r)\Bigl[\frac{\ell(\ell+1)}{r^2}-\frac{6M_S}{r^3}+\frac{24mJ_S(3r-7M_S)}{\ell(\ell+1)\omega r^6}\Bigr] \,  ,
    \end{aligned}
\end{equation}
and $f(r)=1-2M_S/r$. The solution must be regular at the origin, requiring $\psi = a_\psi r^{\ell+1}$, and $Z = a_Z r^{\ell+1}$. Additionally, at the surface of the star, $\psi$ and its radial derivative are continuous, and Israel's junction conditions~\cite{Israel:1966rt} must be satisfied. Since viscosity vanishes at the star's surface, the junction conditions are trivially satisfied. However, regularity of the wave equation for $Z$ enforces the following boundary condition (see discussion in Ref.~\cite{Boyanov:2024jge})
\begin{equation}\label{boundary_condition}
    A_1 \frac{dZ}{dr} + A_2 Z + A_3 \frac{d\psi}{dr} + A_4\psi = 0 \, , 
\end{equation}
where the coefficients $A_i$ are provided in the Supplemental Material. These reduce to Eq.(16) in Ref.~\cite{Boyanov:2024jge} in the non-rotating limit.
We integrate the coupled system~\eqref{wave_eqs} from the origin up to the surface, and ``shoot'' for the value of $a_Z/a_\psi$ that satisfies the boundary condition written above~\eqref{boundary_condition}. We use the \texttt{DifferentialEquations.jl}~\cite{rackauckas2017differentialequations} package to solve the radial ODEs, and the \texttt{NLsolve.jl}~\cite{nlsolve2019} package to ensure the boundary conditions are satisfied. Our code is publicly available in~\cite{web:CoG}, and has been tested against an independent routine written in \texttt{Mathematica}, in addition to reproducing the results of~\cite{Boyanov:2024jge} in the non-rotating limit. 

Asymptotically far away, the solution for $\psi$ can be decomposed as a superposition of incoming and outgoing plane waves,
\begin{equation}
    \psi \xrightarrow{r\to\infty} A_{\rm in}e^{i\omega r_\star} + A_{\rm out} e^{-i\omega r_\star} \, .
\end{equation}
We define the reflectivity of the star as $\mathcal{R}^2(\omega) = |A_{\rm out}/A_{\rm in}|^2$. Superradiant amplification is present whenever $\mathcal{R}^2 > 1$. We integrate the equations until sufficiently far away (typically on the order of $10^4$ wavelengths), and ensure that the numerically extracted reflectivity is stable against changes in the extraction radius. 

\begin{figure}
    \centering
    \includegraphics[width=\columnwidth]{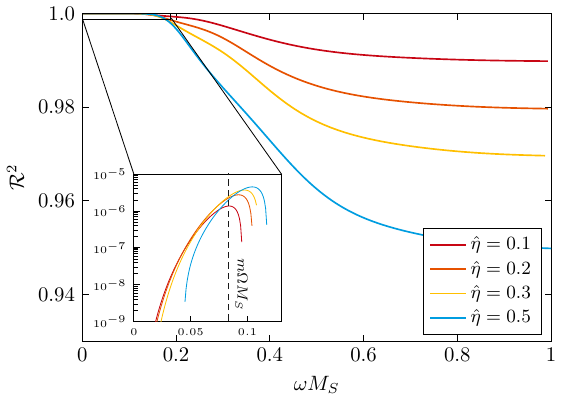}
    \caption{Reflectivity as a function of the dimensionless frequency $\omega M_S$, for different values of the dimensionless shear viscosity parameter $\hat{\eta}$. The inset shows the amplification $\mathcal{R}^2-1$ for frequencies lower than or comparable to the angular frequency of the star. In all three cases we observe superradiant amplification, though the superradiant region varies slightly in each case. We recover similar qualitative results as in~\cite{Boyanov:2024jge} in the high-frequency limit, where viscosity induces absorption. In this case, $\hat{\tau}=500$, and $\Omega/\Omega_K= 0.26$.}
    \label{fig:fig_1}
\end{figure}

\noindent\textbf{\emph{Results.}} 
Our main result is shown in Fig.~\ref{fig:fig_1}: rotating, viscous stars amplify radiation. This is evident in the inset, where the reflectivity exceeds unity, $\mathcal{R}^2>1$, for frequencies $\omega \lesssim m\Omega$. We also find that the maximum amplification increases with the dimensionless shear viscosity. This parameter controls the absorption rate in the high-frequency limit~\cite{Press:1979rd,Boyanov:2024jge}, supporting the idea that stronger absorption leads to shorter superradiant timescales. A similar result was found in Ref.~\cite{Saketh:2024ojw}, in the context of spinning black hole mimickers constructed from the membrane paradigm. The scattering of GWs off a membrane with with a given shear and bulk viscosity also showed the presence of superradiant amplification at low frequencies. However, this effect was classified as spurious in~\cite{Saketh:2024ojw}, because no ergoregions are present at linear order in the spin. As we argue here, superradiance requires no ergoregions -- absorption (induced by viscosity) and rotation provide the necessary mechanism for superradiant amplification.

Amplification is not confined to the classical superradiant regime $0<\omega<m\Omega$, unlike known literature on black holes~\cite{Brito:2015oca}, conducting stars~\cite{Cardoso:2017kgn}, or what one may naively obtain by analyzing simplified models for dissipation (see Ref.~\cite{Cardoso:2015zqa} and Supplemental Material). This is most likely due to the linear-in-spin approximation. The equations of motion include second-order terms in $\omega \lesssim \Omega$, resulting in an inconsistent expansion~\cite{Andersson:1997xt}. In the black hole case this is known to cause amplification for frequencies slightly above the superradiant threshold~\cite{Pani:2012bp}, an effect which disappears once higher-order terms in the rotation rate are included. Non-equilibrium contributions to the entropy balance argument may also alter this bound~\cite{Dias:2007nj}. Numerical instabilities also challenge the extraction of the reflectivity at very low frequencies $\omega \ll m\Omega_K$, where the reflectivity drops below the numerical floor.

\begin{figure}
    \centering
    \includegraphics[width=\columnwidth]{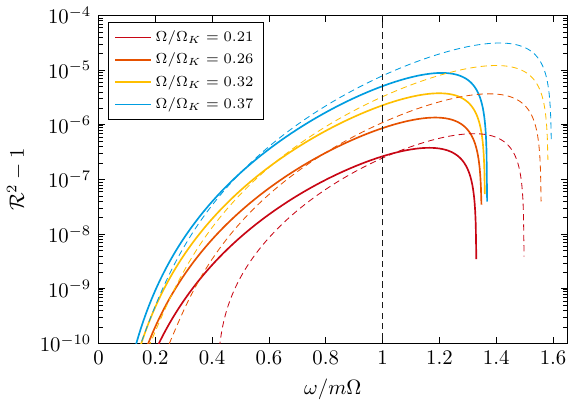}
    \caption{Amplification factor $\mathcal{R}^2-1$ as a function of the dimensionless frequency $\omega/m\Omega$, for different values of the angular velocity of the star $\Omega/\Omega_K$ (see legend). Solid (dashed) lines correspond to $\hat{\eta}=0.1 (0.3)$, while $\hat{\tau}=500$ is fixed. }
    \label{fig:fig_2}
\end{figure}
We investigate the dependence of superradiant amplification on parameter space more thoroughly in Fig.~\ref{fig:fig_2}. Although not shown in the Figure, we report a very mild dependence on $\hat{\tau}$. However, it is particularly challenging to obtain accurate solutions, including superradiance, in the regime where $\hat{\eta}/\hat{\tau}$ is large. 

We can clearly see that superradiant amplification occurs outside the classically allowed region, $\omega > m\Omega$, and this is enhanced by higher values of the dimensionless shear viscosity. We also find that the maximum amplification rate increases both with the rotation rate and the shear viscosity -- higher angular velocities mean more angular momentum is available to be extracted from the star, whereas higher shear viscosity enhances the absorption cross-section~\cite{Boyanov:2024jge}. At low frequencies, the amplification factor becomes very small and lies below the numerical floor.  

\noindent\textbf{\emph{Viscosity-driven instabilities.}} 
Our results could be relevant for the physics of spinning ultracompact objects. When the object is so compact that an unstable light ring is present, a {\it stable} light ring must also feature in the geometry~\cite{Cunha:2017qtt}. Perturbations around such objects decay logarithmically in time~\cite{Keir:2014oka,Cardoso:2014sna,Zhong:2022jke,Benomio:2024lev}. The slow decay of linearized fluctuations has led to the conjecture that nonlinearities may turn such objects unstable~\cite{Keir:2014oka,Cardoso:2014sna}, although recent results fail to find evidence of such instability~\cite{Redondo-Yuste:2025hlv,Marks:2025jpt}. Through numerical examination, we find that the frequency of the modes trapped in ultracompact objects scales as $\omega \sim m (\Omega_{\rm sLR}+\Omega)$, where $\Omega_{\rm sLR}=e^{\nu/2}/r$ is the angular frequency of the stable light ring in the non-rotating limit, and $\Omega$ is the angular speed of the object. By making the object compact enough, the lapse $e^\nu \to 0$ inside the object, so $\Omega_{\rm sLR}$ can be made parametrically small. However, the frame dragging term ensures that the frequency of these trapped modes always exceeds the superradiant threshold, $\omega > m\Omega$. Amplified waves are not trapped, and trapped radiation is not amplified. We have verified this by analyzing the propagation of scalar waves, absorbed in the interior of ultracompact, constant density stars (see Supplemental Material). This supports the linear stability of ultracompact objects, even with rotation and dissipation, though it remains an open question whether an instability could be triggered in the rapidly rotating limit, when ergoregions may appear, or when accounting for nonlinear effects. 

The above concerns uniform rotation. From the above, it is also quite likely that differentially rotating stars may be strongly impacted by viscous instabilities. Indeed, the physics of viscous fluids may also render accretion disks -- intrinsically differentially rotating structures -- dynamically unstable. 

\noindent\textbf{\emph{Discussion.}} 
We have shown that dissipation in stars leads to the amplification of low-frequency gravitational waves, an important step in Zel'dovich's superradiant program. Together with the amplification of any massless field by black holes~\cite{Teukolsky:1974yv,Brito:2015oca}, of sound waves by fluids~\cite{Berti:2004ju,Cardoso:2016zvz} (recently reported experimentally~\cite{Torres:2016iee}), and of
electromagnetic waves by conducting materials~\cite{zeldovich1,zeldovich2,Bekenstein:1998nt} (recently reported experimentally~\cite{Cromb:2025dqu}), our results paint a clear picture of energy extraction by spinning objects.

Our findings for the regime of superradiant amplification are considerably more complex than previously reported in gravitational or electromagnetic wave superradiance. However, we find a similarly rich pattern in the amplification of sound waves at planar interfaces that separate an ideal fluid from a viscous one~\cite{1957ASAJ...29..435R,1954ASAJ...26.1015M}. A follow-up problem is to examine the back-reaction of superradiant amplification onto the star -- and consequent spindown, in the presence of dissipative effects.

Note that our analysis assumes slow rotation, retaining only terms linear in $\epsilon = \Omega / \Omega_K$. However, by examining the regime $\omega \lesssim \Omega$, we effectively incorporate some $\mathcal{O}(\epsilon^2)$ contributions, formally beyond our approximation. This issue is known from early studies of the r-mode instability~\cite{Andersson:1997xt}. Thus, results at high angular velocities or low frequencies should be considered merely informative, serving as a useful baseline. A second-order calculation, though technically challenging, would be a natural next step to better quantify the amplification of radiation by rotating, compact objects.

\textbf{\emph{Acknowledgments.}} 
We are indebted to Caio Macedo for sharing a code to cross-validate some of our numerical results and for insightful comments.
We thank Yifan Chen, Elisa Maggio, and Paolo Pani for valuable comments on this manuscript.
We are grateful to the organizers and participants of the workshop ``Black Hole Mimickers: From Theory to Observation'' at Princeton. The discussions held there renewed our interest in this problem.
The Center of Gravity is a Center of Excellence funded by the Danish National Research Foundation under grant No. 184.
We acknowledge support by VILLUM Foundation (grant no.\ VIL37766) and the DNRF Chair program (grant no.\ DNRF162) by the Danish National Research Foundation.
V.C. acknowledges financial support provided under the European Union’s H2020 ERC Advanced Grant “Black holes: gravitational engines of discovery” grant agreement no.\ Gravitas–101052587. 
Views and opinions expressed are however those of the author only and do not necessarily reflect those of the European Union or the European Research Council. Neither the European Union nor the granting authority can be held responsible for them.
This project has received funding from the European Union's Horizon 2020 research and innovation programme under the Marie Sk{\l}odowska-Curie grant agreement No 101007855 and No 101131233.
The Tycho supercomputer hosted at the SCIENCE HPC center at the University of Copenhagen was used for supporting this work. 

\bibliography{biblio}

\begin{thebibliography}{78}%
\makeatletter
\providecommand \@ifxundefined [1]{%
 \@ifx{#1\undefined}
}%
\providecommand \@ifnum [1]{%
 \ifnum #1\expandafter \@firstoftwo
 \else \expandafter \@secondoftwo
 \fi
}%
\providecommand \@ifx [1]{%
 \ifx #1\expandafter \@firstoftwo
 \else \expandafter \@secondoftwo
 \fi
}%
\providecommand \natexlab [1]{#1}%
\providecommand \enquote  [1]{``#1''}%
\providecommand \bibnamefont  [1]{#1}%
\providecommand \bibfnamefont [1]{#1}%
\providecommand \citenamefont [1]{#1}%
\providecommand \href@noop [0]{\@secondoftwo}%
\providecommand \href [0]{\begingroup \@sanitize@url \@href}%
\providecommand \@href[1]{\@@startlink{#1}\@@href}%
\providecommand \@@href[1]{\endgroup#1\@@endlink}%
\providecommand \@sanitize@url [0]{\catcode `\\12\catcode `\$12\catcode
  `\&12\catcode `\#12\catcode `\^12\catcode `\_12\catcode `\%12\relax}%
\providecommand \@@startlink[1]{}%
\providecommand \@@endlink[0]{}%
\providecommand \url  [0]{\begingroup\@sanitize@url \@url }%
\providecommand \@url [1]{\endgroup\@href {#1}{\urlprefix }}%
\providecommand \urlprefix  [0]{URL }%
\providecommand \Eprint [0]{\href }%
\providecommand \doibase [0]{http://dx.doi.org/}%
\providecommand \selectlanguage [0]{\@gobble}%
\providecommand \bibinfo  [0]{\@secondoftwo}%
\providecommand \bibfield  [0]{\@secondoftwo}%
\providecommand \translation [1]{[#1]}%
\providecommand \BibitemOpen [0]{}%
\providecommand \bibitemStop [0]{}%
\providecommand \bibitemNoStop [0]{.\EOS\space}%
\providecommand \EOS [0]{\spacefactor3000\relax}%
\providecommand \BibitemShut  [1]{\csname bibitem#1\endcsname}%
\let\auto@bib@innerbib\@empty
\bibitem [{\citenamefont {Redondo-Yuste}(2024)}]{Redondo-Yuste:2024vdb}%
  \BibitemOpen
  \bibfield  {author} {\bibinfo {author} {\bibfnamefont {Jaime}\ \bibnamefont
  {Redondo-Yuste}},\ }\bibfield  {title} {\enquote {\bibinfo {title}
  {{Perturbations of relativistic dissipative stars}},}\ }\href@noop {} {\
  (\bibinfo {year} {2024})},\ \Eprint {http://arxiv.org/abs/2411.16841}
  {arXiv:2411.16841 [gr-qc]} \BibitemShut {NoStop}%
\bibitem [{\citenamefont {Boyanov}\ \emph {et~al.}(2024)\citenamefont
  {Boyanov}, \citenamefont {Cardoso}, \citenamefont {Kokkotas},\ and\
  \citenamefont {Redondo-Yuste}}]{Boyanov:2024jge}%
  \BibitemOpen
  \bibfield  {author} {\bibinfo {author} {\bibfnamefont {Valentin}\
  \bibnamefont {Boyanov}}, \bibinfo {author} {\bibfnamefont {Vitor}\
  \bibnamefont {Cardoso}}, \bibinfo {author} {\bibfnamefont {Kostas~D.}\
  \bibnamefont {Kokkotas}}, \ and\ \bibinfo {author} {\bibfnamefont {Jaime}\
  \bibnamefont {Redondo-Yuste}},\ }\bibfield  {title} {\enquote {\bibinfo
  {title} {{The dynamical response of viscous objects to gravitational
  waves}},}\ }\href@noop {} {\  (\bibinfo {year} {2024})},\ \Eprint
  {http://arxiv.org/abs/2411.16861} {arXiv:2411.16861 [gr-qc]} \BibitemShut
  {NoStop}%
\bibitem [{\citenamefont {Shibata}\ and\ \citenamefont
  {Kiuchi}(2017)}]{Shibata:2017xht}%
  \BibitemOpen
  \bibfield  {author} {\bibinfo {author} {\bibfnamefont {Masaru}\ \bibnamefont
  {Shibata}}\ and\ \bibinfo {author} {\bibfnamefont {Kenta}\ \bibnamefont
  {Kiuchi}},\ }\bibfield  {title} {\enquote {\bibinfo {title} {{Gravitational
  waves from remnant massive neutron stars of binary neutron star merger:
  Viscous hydrodynamics effects}},}\ }\href {\doibase
  10.1103/PhysRevD.95.123003} {\bibfield  {journal} {\bibinfo  {journal} {Phys.
  Rev. D}\ }\textbf {\bibinfo {volume} {95}},\ \bibinfo {pages} {123003}
  (\bibinfo {year} {2017})},\ \Eprint {http://arxiv.org/abs/1705.06142}
  {arXiv:1705.06142 [astro-ph.HE]} \BibitemShut {NoStop}%
\bibitem [{\citenamefont {Alford}\ and\ \citenamefont
  {Harris}(2018)}]{Alford:2018lhf}%
  \BibitemOpen
  \bibfield  {author} {\bibinfo {author} {\bibfnamefont {Mark~G.}\ \bibnamefont
  {Alford}}\ and\ \bibinfo {author} {\bibfnamefont {Steven~P.}\ \bibnamefont
  {Harris}},\ }\bibfield  {title} {\enquote {\bibinfo {title} {{Beta
  equilibrium in neutron star mergers}},}\ }\href {\doibase
  10.1103/PhysRevC.98.065806} {\bibfield  {journal} {\bibinfo  {journal} {Phys.
  Rev. C}\ }\textbf {\bibinfo {volume} {98}},\ \bibinfo {pages} {065806}
  (\bibinfo {year} {2018})},\ \Eprint {http://arxiv.org/abs/1803.00662}
  {arXiv:1803.00662 [nucl-th]} \BibitemShut {NoStop}%
\bibitem [{\citenamefont {Alford}\ \emph {et~al.}(2020)\citenamefont {Alford},
  \citenamefont {Harutyunyan},\ and\ \citenamefont
  {Sedrakian}}]{Alford:2020lla}%
  \BibitemOpen
  \bibfield  {author} {\bibinfo {author} {\bibfnamefont {Mark}\ \bibnamefont
  {Alford}}, \bibinfo {author} {\bibfnamefont {Arus}\ \bibnamefont
  {Harutyunyan}}, \ and\ \bibinfo {author} {\bibfnamefont {Armen}\ \bibnamefont
  {Sedrakian}},\ }\bibfield  {title} {\enquote {\bibinfo {title} {{Bulk Viscous
  Damping of Density Oscillations in Neutron Star Mergers}},}\ }\href {\doibase
  10.3390/particles3020034} {\bibfield  {journal} {\bibinfo  {journal}
  {Particles}\ }\textbf {\bibinfo {volume} {3}},\ \bibinfo {pages} {500--517}
  (\bibinfo {year} {2020})},\ \Eprint {http://arxiv.org/abs/2006.07975}
  {arXiv:2006.07975 [nucl-th]} \BibitemShut {NoStop}%
\bibitem [{\citenamefont {Most}\ \emph {et~al.}(2021)\citenamefont {Most},
  \citenamefont {Harris}, \citenamefont {Plumberg}, \citenamefont {Alford},
  \citenamefont {Noronha}, \citenamefont {Noronha-Hostler}, \citenamefont
  {Pretorius}, \citenamefont {Witek},\ and\ \citenamefont
  {Yunes}}]{Most:2021zvc}%
  \BibitemOpen
  \bibfield  {author} {\bibinfo {author} {\bibfnamefont {Elias~R.}\
  \bibnamefont {Most}}, \bibinfo {author} {\bibfnamefont {Steven~P.}\
  \bibnamefont {Harris}}, \bibinfo {author} {\bibfnamefont {Christopher}\
  \bibnamefont {Plumberg}}, \bibinfo {author} {\bibfnamefont {Mark~G.}\
  \bibnamefont {Alford}}, \bibinfo {author} {\bibfnamefont {Jorge}\
  \bibnamefont {Noronha}}, \bibinfo {author} {\bibfnamefont {Jacquelyn}\
  \bibnamefont {Noronha-Hostler}}, \bibinfo {author} {\bibfnamefont {Frans}\
  \bibnamefont {Pretorius}}, \bibinfo {author} {\bibfnamefont {Helvi}\
  \bibnamefont {Witek}}, \ and\ \bibinfo {author} {\bibfnamefont {Nicol\'as}\
  \bibnamefont {Yunes}},\ }\bibfield  {title} {\enquote {\bibinfo {title}
  {{Projecting the likely importance of weak-interaction-driven bulk viscosity
  in neutron star mergers}},}\ }\href {\doibase 10.1093/mnras/stab2793}
  {\bibfield  {journal} {\bibinfo  {journal} {Mon. Not. Roy. Astron. Soc.}\
  }\textbf {\bibinfo {volume} {509}},\ \bibinfo {pages} {1096--1108} (\bibinfo
  {year} {2021})},\ \Eprint {http://arxiv.org/abs/2107.05094} {arXiv:2107.05094
  [astro-ph.HE]} \BibitemShut {NoStop}%
\bibitem [{\citenamefont {Most}\ \emph {et~al.}(2024)\citenamefont {Most},
  \citenamefont {Haber}, \citenamefont {Harris}, \citenamefont {Zhang},
  \citenamefont {Alford},\ and\ \citenamefont {Noronha}}]{Most:2022yhe}%
  \BibitemOpen
  \bibfield  {author} {\bibinfo {author} {\bibfnamefont {Elias~R.}\
  \bibnamefont {Most}}, \bibinfo {author} {\bibfnamefont {Alexander}\
  \bibnamefont {Haber}}, \bibinfo {author} {\bibfnamefont {Steven~P.}\
  \bibnamefont {Harris}}, \bibinfo {author} {\bibfnamefont {Ziyuan}\
  \bibnamefont {Zhang}}, \bibinfo {author} {\bibfnamefont {Mark~G.}\
  \bibnamefont {Alford}}, \ and\ \bibinfo {author} {\bibfnamefont {Jorge}\
  \bibnamefont {Noronha}},\ }\bibfield  {title} {\enquote {\bibinfo {title}
  {{Emergence of Microphysical Bulk Viscosity in Binary Neutron Star Postmerger
  Dynamics}},}\ }\href {\doibase 10.3847/2041-8213/ad454f} {\bibfield
  {journal} {\bibinfo  {journal} {Astrophys. J. Lett.}\ }\textbf {\bibinfo
  {volume} {967}},\ \bibinfo {pages} {L14} (\bibinfo {year} {2024})},\ \Eprint
  {http://arxiv.org/abs/2207.00442} {arXiv:2207.00442 [astro-ph.HE]}
  \BibitemShut {NoStop}%
\bibitem [{\citenamefont {Chabanov}\ and\ \citenamefont
  {Rezzolla}(2025{\natexlab{a}})}]{Chabanov:2023abq}%
  \BibitemOpen
  \bibfield  {author} {\bibinfo {author} {\bibfnamefont {Michail}\ \bibnamefont
  {Chabanov}}\ and\ \bibinfo {author} {\bibfnamefont {Luciano}\ \bibnamefont
  {Rezzolla}},\ }\bibfield  {title} {\enquote {\bibinfo {title} {{Numerical
  modeling of bulk viscosity in neutron stars}},}\ }\href {\doibase
  10.1103/PhysRevD.111.044074} {\bibfield  {journal} {\bibinfo  {journal}
  {Phys. Rev. D}\ }\textbf {\bibinfo {volume} {111}},\ \bibinfo {pages}
  {044074} (\bibinfo {year} {2025}{\natexlab{a}})},\ \Eprint
  {http://arxiv.org/abs/2311.13027} {arXiv:2311.13027 [gr-qc]} \BibitemShut
  {NoStop}%
\bibitem [{\citenamefont {Chabanov}\ and\ \citenamefont
  {Rezzolla}(2025{\natexlab{b}})}]{Chabanov:2023blf}%
  \BibitemOpen
  \bibfield  {author} {\bibinfo {author} {\bibfnamefont {Michail}\ \bibnamefont
  {Chabanov}}\ and\ \bibinfo {author} {\bibfnamefont {Luciano}\ \bibnamefont
  {Rezzolla}},\ }\bibfield  {title} {\enquote {\bibinfo {title} {{Impact of
  Bulk Viscosity on the Postmerger Gravitational-Wave Signal from Merging
  Neutron Stars}},}\ }\href {\doibase 10.1103/PhysRevLett.134.071402}
  {\bibfield  {journal} {\bibinfo  {journal} {Phys. Rev. Lett.}\ }\textbf
  {\bibinfo {volume} {134}},\ \bibinfo {pages} {071402} (\bibinfo {year}
  {2025}{\natexlab{b}})},\ \Eprint {http://arxiv.org/abs/2307.10464}
  {arXiv:2307.10464 [gr-qc]} \BibitemShut {NoStop}%
\bibitem [{\citenamefont {Chabanov}\ \emph {et~al.}(2024)\citenamefont
  {Chabanov}, \citenamefont {Cruz-Osorio}, \citenamefont {Ecker}, \citenamefont
  {Meringolo}, \citenamefont {Musolino}, \citenamefont {Rezzolla},
  \citenamefont {Tootle},\ and\ \citenamefont {Topolski}}]{Chabanov:2024yqv}%
  \BibitemOpen
  \bibfield  {author} {\bibinfo {author} {\bibfnamefont {Michail}\ \bibnamefont
  {Chabanov}}, \bibinfo {author} {\bibfnamefont {Alejandro}\ \bibnamefont
  {Cruz-Osorio}}, \bibinfo {author} {\bibfnamefont {Christian}\ \bibnamefont
  {Ecker}}, \bibinfo {author} {\bibfnamefont {Claudio}\ \bibnamefont
  {Meringolo}}, \bibinfo {author} {\bibfnamefont {Carlo}\ \bibnamefont
  {Musolino}}, \bibinfo {author} {\bibfnamefont {Luciano}\ \bibnamefont
  {Rezzolla}}, \bibinfo {author} {\bibfnamefont {Samuel}\ \bibnamefont
  {Tootle}}, \ and\ \bibinfo {author} {\bibfnamefont {Konrad}\ \bibnamefont
  {Topolski}},\ }\bibfield  {title} {\enquote {\bibinfo {title} {{Microphysical
  Aspects of Binary Neutron Star Mergers}},}\ }in\ \href {\doibase
  10.1007/978-3-031-46870-4_2} {\emph {\bibinfo {booktitle} {{High Performance
  Computing in Science and Engineering '22}}}}\ (\bibinfo {year}
  {2024})\BibitemShut {NoStop}%
\bibitem [{\citenamefont {Pandya}\ \emph
  {et~al.}(2022{\natexlab{a}})\citenamefont {Pandya}, \citenamefont {Most},\
  and\ \citenamefont {Pretorius}}]{Pandya:2022pif}%
  \BibitemOpen
  \bibfield  {author} {\bibinfo {author} {\bibfnamefont {Alex}\ \bibnamefont
  {Pandya}}, \bibinfo {author} {\bibfnamefont {Elias~R.}\ \bibnamefont {Most}},
  \ and\ \bibinfo {author} {\bibfnamefont {Frans}\ \bibnamefont {Pretorius}},\
  }\bibfield  {title} {\enquote {\bibinfo {title} {{Conservative finite volume
  scheme for first-order viscous relativistic hydrodynamics}},}\ }\href
  {\doibase 10.1103/PhysRevD.105.123001} {\bibfield  {journal} {\bibinfo
  {journal} {Phys. Rev. D}\ }\textbf {\bibinfo {volume} {105}},\ \bibinfo
  {pages} {123001} (\bibinfo {year} {2022}{\natexlab{a}})},\ \Eprint
  {http://arxiv.org/abs/2201.12317} {arXiv:2201.12317 [gr-qc]} \BibitemShut
  {NoStop}%
\bibitem [{\citenamefont {Pandya}\ \emph
  {et~al.}(2022{\natexlab{b}})\citenamefont {Pandya}, \citenamefont {Most},\
  and\ \citenamefont {Pretorius}}]{Pandya:2022sff}%
  \BibitemOpen
  \bibfield  {author} {\bibinfo {author} {\bibfnamefont {Alex}\ \bibnamefont
  {Pandya}}, \bibinfo {author} {\bibfnamefont {Elias~R.}\ \bibnamefont {Most}},
  \ and\ \bibinfo {author} {\bibfnamefont {Frans}\ \bibnamefont {Pretorius}},\
  }\bibfield  {title} {\enquote {\bibinfo {title} {{Causal, stable first-order
  viscous relativistic hydrodynamics with ideal gas microphysics}},}\ }\href
  {\doibase 10.1103/PhysRevD.106.123036} {\bibfield  {journal} {\bibinfo
  {journal} {Phys. Rev. D}\ }\textbf {\bibinfo {volume} {106}},\ \bibinfo
  {pages} {123036} (\bibinfo {year} {2022}{\natexlab{b}})},\ \Eprint
  {http://arxiv.org/abs/2209.09265} {arXiv:2209.09265 [gr-qc]} \BibitemShut
  {NoStop}%
\bibitem [{\citenamefont {Ripley}\ \emph {et~al.}(2024)\citenamefont {Ripley},
  \citenamefont {Hegade K.~R.}, \citenamefont {Chandramouli},\ and\
  \citenamefont {Yunes}}]{Ripley:2023lsq}%
  \BibitemOpen
  \bibfield  {author} {\bibinfo {author} {\bibfnamefont {Justin~L.}\
  \bibnamefont {Ripley}}, \bibinfo {author} {\bibfnamefont {Abhishek}\
  \bibnamefont {Hegade K.~R.}}, \bibinfo {author} {\bibfnamefont {Rohit~S.}\
  \bibnamefont {Chandramouli}}, \ and\ \bibinfo {author} {\bibfnamefont
  {Nicolas}\ \bibnamefont {Yunes}},\ }\bibfield  {title} {\enquote {\bibinfo
  {title} {{A constraint on the dissipative tidal deformability of neutron
  stars}},}\ }\href {\doibase 10.1038/s41550-024-02323-7} {\bibfield  {journal}
  {\bibinfo  {journal} {Nature Astron.}\ }\textbf {\bibinfo {volume} {8}},\
  \bibinfo {pages} {1277--1283} (\bibinfo {year} {2024})},\ \Eprint
  {http://arxiv.org/abs/2312.11659} {arXiv:2312.11659 [gr-qc]} \BibitemShut
  {NoStop}%
\bibitem [{\citenamefont {Ripley}\ \emph {et~al.}(2023)\citenamefont {Ripley},
  \citenamefont {Hegade K.~R.},\ and\ \citenamefont {Yunes}}]{Ripley:2023qxo}%
  \BibitemOpen
  \bibfield  {author} {\bibinfo {author} {\bibfnamefont {Justin~L.}\
  \bibnamefont {Ripley}}, \bibinfo {author} {\bibfnamefont {Abhishek}\
  \bibnamefont {Hegade K.~R.}}, \ and\ \bibinfo {author} {\bibfnamefont
  {Nicolas}\ \bibnamefont {Yunes}},\ }\bibfield  {title} {\enquote {\bibinfo
  {title} {{Probing internal dissipative processes of neutron stars with
  gravitational waves during the inspiral of neutron star binaries}},}\ }\href
  {\doibase 10.1103/PhysRevD.108.103037} {\bibfield  {journal} {\bibinfo
  {journal} {Phys. Rev. D}\ }\textbf {\bibinfo {volume} {108}},\ \bibinfo
  {pages} {103037} (\bibinfo {year} {2023})},\ \Eprint
  {http://arxiv.org/abs/2306.15633} {arXiv:2306.15633 [gr-qc]} \BibitemShut
  {NoStop}%
\bibitem [{\citenamefont {Hegade K.~R.}\ \emph
  {et~al.}(2024{\natexlab{a}})\citenamefont {Hegade K.~R.}, \citenamefont
  {Ripley},\ and\ \citenamefont {Yunes}}]{HegadeKR:2024agt}%
  \BibitemOpen
  \bibfield  {author} {\bibinfo {author} {\bibfnamefont {Abhishek}\
  \bibnamefont {Hegade K.~R.}}, \bibinfo {author} {\bibfnamefont {Justin~L.}\
  \bibnamefont {Ripley}}, \ and\ \bibinfo {author} {\bibfnamefont {Nicol\'as}\
  \bibnamefont {Yunes}},\ }\bibfield  {title} {\enquote {\bibinfo {title}
  {{Dynamical tidal response of nonrotating relativistic stars}},}\ }\href
  {\doibase 10.1103/PhysRevD.109.104064} {\bibfield  {journal} {\bibinfo
  {journal} {Phys. Rev. D}\ }\textbf {\bibinfo {volume} {109}},\ \bibinfo
  {pages} {104064} (\bibinfo {year} {2024}{\natexlab{a}})},\ \Eprint
  {http://arxiv.org/abs/2403.03254} {arXiv:2403.03254 [gr-qc]} \BibitemShut
  {NoStop}%
\bibitem [{\citenamefont {Hegade K.~R.}\ \emph
  {et~al.}(2024{\natexlab{b}})\citenamefont {Hegade K.~R.}, \citenamefont
  {Ripley},\ and\ \citenamefont {Yunes}}]{HegadeKR:2024slr}%
  \BibitemOpen
  \bibfield  {author} {\bibinfo {author} {\bibfnamefont {Abhishek}\
  \bibnamefont {Hegade K.~R.}}, \bibinfo {author} {\bibfnamefont {Justin~L.}\
  \bibnamefont {Ripley}}, \ and\ \bibinfo {author} {\bibfnamefont {Nicol\'as}\
  \bibnamefont {Yunes}},\ }\bibfield  {title} {\enquote {\bibinfo {title}
  {{Dissipative tidal effects to next-to-leading order and constraints on the
  dissipative tidal deformability using gravitational wave data}},}\ }\href
  {\doibase 10.1103/PhysRevD.110.044041} {\bibfield  {journal} {\bibinfo
  {journal} {Phys. Rev. D}\ }\textbf {\bibinfo {volume} {110}},\ \bibinfo
  {pages} {044041} (\bibinfo {year} {2024}{\natexlab{b}})},\ \Eprint
  {http://arxiv.org/abs/2407.02584} {arXiv:2407.02584 [gr-qc]} \BibitemShut
  {NoStop}%
\bibitem [{\citenamefont {Caballero}\ and\ \citenamefont
  {Yunes}(2025)}]{Caballero:2025omv}%
  \BibitemOpen
  \bibfield  {author} {\bibinfo {author} {\bibfnamefont {Daniel~A.}\
  \bibnamefont {Caballero}}\ and\ \bibinfo {author} {\bibfnamefont {Nicol\'as}\
  \bibnamefont {Yunes}},\ }\bibfield  {title} {\enquote {\bibinfo {title}
  {{Neutron Star Radial Perturbations for Causal, Viscous, Relativistic
  Fluids}},}\ }\href@noop {} {\  (\bibinfo {year} {2025})},\ \Eprint
  {http://arxiv.org/abs/2506.09149} {arXiv:2506.09149 [gr-qc]} \BibitemShut
  {NoStop}%
\bibitem [{\citenamefont {Cardoso}\ and\ \citenamefont
  {Pani}(2019)}]{Cardoso:2019rvt}%
  \BibitemOpen
  \bibfield  {author} {\bibinfo {author} {\bibfnamefont {Vitor}\ \bibnamefont
  {Cardoso}}\ and\ \bibinfo {author} {\bibfnamefont {Paolo}\ \bibnamefont
  {Pani}},\ }\bibfield  {title} {\enquote {\bibinfo {title} {{Testing the
  nature of dark compact objects: a status report}},}\ }\href {\doibase
  10.1007/s41114-019-0020-4} {\bibfield  {journal} {\bibinfo  {journal} {Living
  Rev. Rel.}\ }\textbf {\bibinfo {volume} {22}},\ \bibinfo {pages} {4}
  (\bibinfo {year} {2019})},\ \Eprint {http://arxiv.org/abs/1904.05363}
  {arXiv:1904.05363 [gr-qc]} \BibitemShut {NoStop}%
\bibitem [{\citenamefont {Afshordi}\ \emph {et~al.}(2024)\citenamefont
  {Afshordi} \emph {et~al.}}]{Buoninfante:2024oxl}%
  \BibitemOpen
  \bibfield  {author} {\bibinfo {author} {\bibfnamefont {Niayesh}\ \bibnamefont
  {Afshordi}} \emph {et~al.},\ }\bibfield  {title} {\enquote {\bibinfo {title}
  {{Black Holes Inside and Out 2024: visions for the future of black hole
  physics}},}\ \ }(\bibinfo {year} {2024})\ \Eprint
  {http://arxiv.org/abs/2410.14414} {arXiv:2410.14414 [gr-qc]} \BibitemShut
  {NoStop}%
\bibitem [{\citenamefont {Bambi}\ \emph {et~al.}(2025)\citenamefont {Bambi}
  \emph {et~al.}}]{Bambi:2025wjx}%
  \BibitemOpen
  \bibfield  {author} {\bibinfo {author} {\bibfnamefont {Cosimo}\ \bibnamefont
  {Bambi}} \emph {et~al.},\ }\bibfield  {title} {\enquote {\bibinfo {title}
  {{Black hole mimickers: from theory to observation}},}\ \ }(\bibinfo {year}
  {2025})\ \Eprint {http://arxiv.org/abs/2505.09014} {arXiv:2505.09014 [gr-qc]}
  \BibitemShut {NoStop}%
\bibitem [{\citenamefont {Friedman}(1978)}]{Friedman:1978ygc}%
  \BibitemOpen
  \bibfield  {author} {\bibinfo {author} {\bibfnamefont {John~L.}\ \bibnamefont
  {Friedman}},\ }\bibfield  {title} {\enquote {\bibinfo {title} {{Ergosphere
  instability}},}\ }\href {\doibase 10.1007/BF01196933} {\bibfield  {journal}
  {\bibinfo  {journal} {Commun. Math. Phys.}\ }\textbf {\bibinfo {volume}
  {63}},\ \bibinfo {pages} {243--255} (\bibinfo {year} {1978})}\BibitemShut
  {NoStop}%
\bibitem [{\citenamefont {Moschidis}(2018)}]{Moschidis:2016zjy}%
  \BibitemOpen
  \bibfield  {author} {\bibinfo {author} {\bibfnamefont {Georgios}\
  \bibnamefont {Moschidis}},\ }\bibfield  {title} {\enquote {\bibinfo {title}
  {{A Proof of Friedman\textquoteright{}s Ergosphere Instability for Scalar
  Waves}},}\ }\href {\doibase 10.1007/s00220-017-3010-y} {\bibfield  {journal}
  {\bibinfo  {journal} {Commun. Math. Phys.}\ }\textbf {\bibinfo {volume}
  {358}},\ \bibinfo {pages} {437--520} (\bibinfo {year} {2018})},\ \Eprint
  {http://arxiv.org/abs/1608.02035} {arXiv:1608.02035 [math.AP]} \BibitemShut
  {NoStop}%
\bibitem [{\citenamefont {Barausse}\ \emph {et~al.}(2018)\citenamefont
  {Barausse}, \citenamefont {Brito}, \citenamefont {Cardoso}, \citenamefont
  {Dvorkin},\ and\ \citenamefont {Pani}}]{Barausse:2018vdb}%
  \BibitemOpen
  \bibfield  {author} {\bibinfo {author} {\bibfnamefont {Enrico}\ \bibnamefont
  {Barausse}}, \bibinfo {author} {\bibfnamefont {Richard}\ \bibnamefont
  {Brito}}, \bibinfo {author} {\bibfnamefont {Vitor}\ \bibnamefont {Cardoso}},
  \bibinfo {author} {\bibfnamefont {Irina}\ \bibnamefont {Dvorkin}}, \ and\
  \bibinfo {author} {\bibfnamefont {Paolo}\ \bibnamefont {Pani}},\ }\bibfield
  {title} {\enquote {\bibinfo {title} {{The stochastic gravitational-wave
  background in the absence of horizons}},}\ }\href {\doibase
  10.1088/1361-6382/aae1de} {\bibfield  {journal} {\bibinfo  {journal} {Class.
  Quant. Grav.}\ }\textbf {\bibinfo {volume} {35}},\ \bibinfo {pages} {20LT01}
  (\bibinfo {year} {2018})},\ \Eprint {http://arxiv.org/abs/1805.08229}
  {arXiv:1805.08229 [gr-qc]} \BibitemShut {NoStop}%
\bibitem [{\citenamefont {Brito}\ \emph {et~al.}(2015)\citenamefont {Brito},
  \citenamefont {Cardoso},\ and\ \citenamefont {Pani}}]{Brito:2015oca}%
  \BibitemOpen
  \bibfield  {author} {\bibinfo {author} {\bibfnamefont {Richard}\ \bibnamefont
  {Brito}}, \bibinfo {author} {\bibfnamefont {Vitor}\ \bibnamefont {Cardoso}},
  \ and\ \bibinfo {author} {\bibfnamefont {Paolo}\ \bibnamefont {Pani}},\
  }\bibfield  {title} {\enquote {\bibinfo {title} {{Superradiance}: {New
  Frontiers in Black Hole Physics}},}\ }\href {\doibase
  10.1007/978-3-319-19000-6} {\bibfield  {journal} {\bibinfo  {journal} {Lect.
  Notes Phys.}\ }\textbf {\bibinfo {volume} {906}},\ \bibinfo {pages}
  {pp.1--237} (\bibinfo {year} {2015})},\ \Eprint
  {http://arxiv.org/abs/1501.06570} {arXiv:1501.06570 [gr-qc]} \BibitemShut
  {NoStop}%
\bibitem [{\citenamefont {Andersson}(1998)}]{Andersson:1997xt}%
  \BibitemOpen
  \bibfield  {author} {\bibinfo {author} {\bibfnamefont {Nils}\ \bibnamefont
  {Andersson}},\ }\bibfield  {title} {\enquote {\bibinfo {title} {{A New class
  of unstable modes of rotating relativistic stars}},}\ }\href {\doibase
  10.1086/305919} {\bibfield  {journal} {\bibinfo  {journal} {Astrophys. J.}\
  }\textbf {\bibinfo {volume} {502}},\ \bibinfo {pages} {708--713} (\bibinfo
  {year} {1998})},\ \Eprint {http://arxiv.org/abs/gr-qc/9706075}
  {arXiv:gr-qc/9706075} \BibitemShut {NoStop}%
\bibitem [{\citenamefont {Lindblom}\ \emph {et~al.}(1998)\citenamefont
  {Lindblom}, \citenamefont {Owen},\ and\ \citenamefont
  {Morsink}}]{Lindblom:1998wf}%
  \BibitemOpen
  \bibfield  {author} {\bibinfo {author} {\bibfnamefont {Lee}\ \bibnamefont
  {Lindblom}}, \bibinfo {author} {\bibfnamefont {Benjamin~J.}\ \bibnamefont
  {Owen}}, \ and\ \bibinfo {author} {\bibfnamefont {Sharon~M.}\ \bibnamefont
  {Morsink}},\ }\bibfield  {title} {\enquote {\bibinfo {title} {{Gravitational
  radiation instability in hot young neutron stars}},}\ }\href {\doibase
  10.1103/PhysRevLett.80.4843} {\bibfield  {journal} {\bibinfo  {journal}
  {Phys. Rev. Lett.}\ }\textbf {\bibinfo {volume} {80}},\ \bibinfo {pages}
  {4843--4846} (\bibinfo {year} {1998})},\ \Eprint
  {http://arxiv.org/abs/gr-qc/9803053} {arXiv:gr-qc/9803053} \BibitemShut
  {NoStop}%
\bibitem [{\citenamefont {Andersson}\ and\ \citenamefont
  {Kokkotas}(2001)}]{Andersson:2000mf}%
  \BibitemOpen
  \bibfield  {author} {\bibinfo {author} {\bibfnamefont {Nils}\ \bibnamefont
  {Andersson}}\ and\ \bibinfo {author} {\bibfnamefont {Kostas~D.}\ \bibnamefont
  {Kokkotas}},\ }\bibfield  {title} {\enquote {\bibinfo {title} {{The R mode
  instability in rotating neutron stars}},}\ }\href {\doibase
  10.1142/S0218271801001062} {\bibfield  {journal} {\bibinfo  {journal} {Int.
  J. Mod. Phys. D}\ }\textbf {\bibinfo {volume} {10}},\ \bibinfo {pages}
  {381--442} (\bibinfo {year} {2001})},\ \Eprint
  {http://arxiv.org/abs/gr-qc/0010102} {arXiv:gr-qc/0010102} \BibitemShut
  {NoStop}%
\bibitem [{\citenamefont {Lindblom}\ \emph {et~al.}(1999)\citenamefont
  {Lindblom}, \citenamefont {Mendell},\ and\ \citenamefont
  {Owen}}]{Lindblom:1999yk}%
  \BibitemOpen
  \bibfield  {author} {\bibinfo {author} {\bibfnamefont {Lee}\ \bibnamefont
  {Lindblom}}, \bibinfo {author} {\bibfnamefont {Gregory}\ \bibnamefont
  {Mendell}}, \ and\ \bibinfo {author} {\bibfnamefont {Benjamin~J.}\
  \bibnamefont {Owen}},\ }\bibfield  {title} {\enquote {\bibinfo {title}
  {{Second order rotational effects on the r modes of neutron stars}},}\ }\href
  {\doibase 10.1103/PhysRevD.60.064006} {\bibfield  {journal} {\bibinfo
  {journal} {Phys. Rev. D}\ }\textbf {\bibinfo {volume} {60}},\ \bibinfo
  {pages} {064006} (\bibinfo {year} {1999})},\ \Eprint
  {http://arxiv.org/abs/gr-qc/9902052} {arXiv:gr-qc/9902052} \BibitemShut
  {NoStop}%
\bibitem [{\citenamefont {Lindblom}\ \emph {et~al.}(2000)\citenamefont
  {Lindblom}, \citenamefont {Owen},\ and\ \citenamefont
  {Ushomirsky}}]{Lindblom:2000gu}%
  \BibitemOpen
  \bibfield  {author} {\bibinfo {author} {\bibfnamefont {Lee}\ \bibnamefont
  {Lindblom}}, \bibinfo {author} {\bibfnamefont {Benjamin~J.}\ \bibnamefont
  {Owen}}, \ and\ \bibinfo {author} {\bibfnamefont {Greg}\ \bibnamefont
  {Ushomirsky}},\ }\bibfield  {title} {\enquote {\bibinfo {title} {{Effect of a
  neutron star crust on the r mode instability}},}\ }\href {\doibase
  10.1103/PhysRevD.62.084030} {\bibfield  {journal} {\bibinfo  {journal} {Phys.
  Rev. D}\ }\textbf {\bibinfo {volume} {62}},\ \bibinfo {pages} {084030}
  (\bibinfo {year} {2000})},\ \Eprint {http://arxiv.org/abs/astro-ph/0006242}
  {arXiv:astro-ph/0006242} \BibitemShut {NoStop}%
\bibitem [{\citenamefont {Owen}\ \emph {et~al.}(1998)\citenamefont {Owen},
  \citenamefont {Lindblom}, \citenamefont {Cutler}, \citenamefont {Schutz},
  \citenamefont {Vecchio},\ and\ \citenamefont {Andersson}}]{Owen:1998xg}%
  \BibitemOpen
  \bibfield  {author} {\bibinfo {author} {\bibfnamefont {Benjamin~J.}\
  \bibnamefont {Owen}}, \bibinfo {author} {\bibfnamefont {Lee}\ \bibnamefont
  {Lindblom}}, \bibinfo {author} {\bibfnamefont {Curt}\ \bibnamefont {Cutler}},
  \bibinfo {author} {\bibfnamefont {Bernard~F.}\ \bibnamefont {Schutz}},
  \bibinfo {author} {\bibfnamefont {Alberto}\ \bibnamefont {Vecchio}}, \ and\
  \bibinfo {author} {\bibfnamefont {Nils}\ \bibnamefont {Andersson}},\
  }\bibfield  {title} {\enquote {\bibinfo {title} {{Gravitational waves from
  hot young rapidly rotating neutron stars}},}\ }\href {\doibase
  10.1103/PhysRevD.58.084020} {\bibfield  {journal} {\bibinfo  {journal} {Phys.
  Rev. D}\ }\textbf {\bibinfo {volume} {58}},\ \bibinfo {pages} {084020}
  (\bibinfo {year} {1998})},\ \Eprint {http://arxiv.org/abs/gr-qc/9804044}
  {arXiv:gr-qc/9804044} \BibitemShut {NoStop}%
\bibitem [{\citenamefont {Lindblom}\ and\ \citenamefont
  {Owen}(2002)}]{Lindblom:2001hd}%
  \BibitemOpen
  \bibfield  {author} {\bibinfo {author} {\bibfnamefont {Lee}\ \bibnamefont
  {Lindblom}}\ and\ \bibinfo {author} {\bibfnamefont {Benjamin~J.}\
  \bibnamefont {Owen}},\ }\bibfield  {title} {\enquote {\bibinfo {title}
  {{Effect of hyperon bulk viscosity on neutron star r modes}},}\ }\href
  {\doibase 10.1103/PhysRevD.65.063006} {\bibfield  {journal} {\bibinfo
  {journal} {Phys. Rev. D}\ }\textbf {\bibinfo {volume} {65}},\ \bibinfo
  {pages} {063006} (\bibinfo {year} {2002})},\ \Eprint
  {http://arxiv.org/abs/astro-ph/0110558} {arXiv:astro-ph/0110558} \BibitemShut
  {NoStop}%
\bibitem [{\citenamefont {Bildsten}\ and\ \citenamefont
  {Ushomirsky}(2000)}]{Bildsten:1999zn}%
  \BibitemOpen
  \bibfield  {author} {\bibinfo {author} {\bibfnamefont {Lars}\ \bibnamefont
  {Bildsten}}\ and\ \bibinfo {author} {\bibfnamefont {Greg}\ \bibnamefont
  {Ushomirsky}},\ }\bibfield  {title} {\enquote {\bibinfo {title} {{Viscous
  boundary layer damping of R modes in neutron stars}},}\ }\href {\doibase
  10.1086/312454} {\bibfield  {journal} {\bibinfo  {journal} {Astrophys. J.
  Lett.}\ }\textbf {\bibinfo {volume} {529}},\ \bibinfo {pages} {L33--L36}
  (\bibinfo {year} {2000})},\ \Eprint {http://arxiv.org/abs/astro-ph/9911155}
  {arXiv:astro-ph/9911155} \BibitemShut {NoStop}%
\bibitem [{\citenamefont {Kraav}\ \emph {et~al.}(2024)\citenamefont {Kraav},
  \citenamefont {Gusakov},\ and\ \citenamefont {Kantor}}]{Kraav:2024cus}%
  \BibitemOpen
  \bibfield  {author} {\bibinfo {author} {\bibfnamefont {Kirill~Y.}\
  \bibnamefont {Kraav}}, \bibinfo {author} {\bibfnamefont {Mikhail~E.}\
  \bibnamefont {Gusakov}}, \ and\ \bibinfo {author} {\bibfnamefont {Elena~M.}\
  \bibnamefont {Kantor}},\ }\bibfield  {title} {\enquote {\bibinfo {title}
  {{Instability windows of relativistic r-modes}},}\ }\href {\doibase
  10.1103/PhysRevD.109.043012} {\bibfield  {journal} {\bibinfo  {journal}
  {Phys. Rev. D}\ }\textbf {\bibinfo {volume} {109}},\ \bibinfo {pages}
  {043012} (\bibinfo {year} {2024})},\ \Eprint
  {http://arxiv.org/abs/2401.06200} {arXiv:2401.06200 [astro-ph.HE]}
  \BibitemShut {NoStop}%
\bibitem [{\citenamefont {Pons}\ \emph {et~al.}(2005)\citenamefont {Pons},
  \citenamefont {Gualtieri}, \citenamefont {Miralles},\ and\ \citenamefont
  {Ferrari}}]{Pons:2005gb}%
  \BibitemOpen
  \bibfield  {author} {\bibinfo {author} {\bibfnamefont {J.~A.}\ \bibnamefont
  {Pons}}, \bibinfo {author} {\bibfnamefont {L.}~\bibnamefont {Gualtieri}},
  \bibinfo {author} {\bibfnamefont {J.~A.}\ \bibnamefont {Miralles}}, \ and\
  \bibinfo {author} {\bibfnamefont {V.}~\bibnamefont {Ferrari}},\ }\bibfield
  {title} {\enquote {\bibinfo {title} {{Relativistic r-modes and shear
  viscosity: Regularizing the continuous spectrum}},}\ }\href {\doibase
  10.1111/j.1365-2966.2005.09429.x} {\bibfield  {journal} {\bibinfo  {journal}
  {Mon. Not. Roy. Astron. Soc.}\ }\textbf {\bibinfo {volume} {363}},\ \bibinfo
  {pages} {121--130} (\bibinfo {year} {2005})},\ \Eprint
  {http://arxiv.org/abs/astro-ph/0504062} {arXiv:astro-ph/0504062} \BibitemShut
  {NoStop}%
\bibitem [{\citenamefont {{Ribner}}(1957)}]{1957ASAJ...29..435R}%
  \BibitemOpen
  \bibfield  {author} {\bibinfo {author} {\bibfnamefont {Herbert~S.}\
  \bibnamefont {{Ribner}}},\ }\bibfield  {title} {\enquote {\bibinfo {title}
  {{Reflection, Transmission, and Amplification of Sound by a Moving
  Medium}},}\ }\href {\doibase 10.1121/1.1908918} {\bibfield  {journal}
  {\bibinfo  {journal} {Acoustical Society of America Journal}\ }\textbf
  {\bibinfo {volume} {29}},\ \bibinfo {pages} {435} (\bibinfo {year}
  {1957})}\BibitemShut {NoStop}%
\bibitem [{\citenamefont {{Miles}}(1954)}]{1954ASAJ...26.1015M}%
  \BibitemOpen
  \bibfield  {author} {\bibinfo {author} {\bibfnamefont {John~W.}\ \bibnamefont
  {{Miles}}},\ }\bibfield  {title} {\enquote {\bibinfo {title} {{Dispersive
  Reflection at the Interface between Ideal and Viscous Media}},}\ }\href
  {\doibase 10.1121/1.1907439} {\bibfield  {journal} {\bibinfo  {journal}
  {Acoustical Society of America Journal}\ }\textbf {\bibinfo {volume} {26}},\
  \bibinfo {pages} {1015} (\bibinfo {year} {1954})}\BibitemShut {NoStop}%
\bibitem [{\citenamefont {Zel'dovich}(1971)}]{zeldovich1}%
  \BibitemOpen
  \bibfield  {author} {\bibinfo {author} {\bibfnamefont {Ya.~B.}\ \bibnamefont
  {Zel'dovich}},\ }\href@noop {} {\bibfield  {journal} {\bibinfo  {journal}
  {Pis'ma Zh. Eksp. Teor. Fiz.}\ }\textbf {\bibinfo {volume} {14}},\ \bibinfo
  {pages} {270 [JETP Lett. {\bf14}, 180 (1971)]} (\bibinfo {year}
  {1971})}\BibitemShut {NoStop}%
\bibitem [{\citenamefont {Zel'dovich}(1972)}]{zeldovich2}%
  \BibitemOpen
  \bibfield  {author} {\bibinfo {author} {\bibfnamefont {Ya.~B.}\ \bibnamefont
  {Zel'dovich}},\ }\href@noop {} {\bibfield  {journal} {\bibinfo  {journal}
  {Zh. Eksp. Teor. Fiz}\ }\textbf {\bibinfo {volume} {62}},\ \bibinfo {pages}
  {2076 [Sov.Phys. JETP {\bf 35}, 1085 (1972)]} (\bibinfo {year}
  {1972})}\BibitemShut {NoStop}%
\bibitem [{\citenamefont {Bemfica}\ \emph {et~al.}(2018)\citenamefont
  {Bemfica}, \citenamefont {Disconzi},\ and\ \citenamefont
  {Noronha}}]{Bemfica:2017wps}%
  \BibitemOpen
  \bibfield  {author} {\bibinfo {author} {\bibfnamefont {F\'abio~S.}\
  \bibnamefont {Bemfica}}, \bibinfo {author} {\bibfnamefont {Marcelo~M.}\
  \bibnamefont {Disconzi}}, \ and\ \bibinfo {author} {\bibfnamefont {Jorge}\
  \bibnamefont {Noronha}},\ }\bibfield  {title} {\enquote {\bibinfo {title}
  {{Causality and existence of solutions of relativistic viscous fluid dynamics
  with gravity}},}\ }\href {\doibase 10.1103/PhysRevD.98.104064} {\bibfield
  {journal} {\bibinfo  {journal} {Phys. Rev. D}\ }\textbf {\bibinfo {volume}
  {98}},\ \bibinfo {pages} {104064} (\bibinfo {year} {2018})},\ \Eprint
  {http://arxiv.org/abs/1708.06255} {arXiv:1708.06255 [gr-qc]} \BibitemShut
  {NoStop}%
\bibitem [{\citenamefont {Bemfica}\ \emph {et~al.}(2019)\citenamefont
  {Bemfica}, \citenamefont {Bemfica}, \citenamefont {Disconzi}, \citenamefont
  {Disconzi}, \citenamefont {Noronha},\ and\ \citenamefont
  {Noronha}}]{Bemfica:2019knx}%
  \BibitemOpen
  \bibfield  {author} {\bibinfo {author} {\bibfnamefont {F\'abio~S.}\
  \bibnamefont {Bemfica}}, \bibinfo {author} {\bibfnamefont {F\'abio~S.}\
  \bibnamefont {Bemfica}}, \bibinfo {author} {\bibfnamefont {Marcelo~M.}\
  \bibnamefont {Disconzi}}, \bibinfo {author} {\bibfnamefont {Marcelo~M.}\
  \bibnamefont {Disconzi}}, \bibinfo {author} {\bibfnamefont {Jorge}\
  \bibnamefont {Noronha}}, \ and\ \bibinfo {author} {\bibfnamefont {Jorge}\
  \bibnamefont {Noronha}},\ }\bibfield  {title} {\enquote {\bibinfo {title}
  {{Nonlinear Causality of General First-Order Relativistic Viscous
  Hydrodynamics}},}\ }\href {\doibase 10.1103/PhysRevD.100.104020} {\bibfield
  {journal} {\bibinfo  {journal} {Phys. Rev. D}\ }\textbf {\bibinfo {volume}
  {100}},\ \bibinfo {pages} {104020} (\bibinfo {year} {2019})},\ \bibinfo
  {note} {[Erratum: Phys.Rev.D 105, 069902 (2022)]},\ \Eprint
  {http://arxiv.org/abs/1907.12695} {arXiv:1907.12695 [gr-qc]} \BibitemShut
  {NoStop}%
\bibitem [{\citenamefont {Bemfica}\ \emph {et~al.}(2022)\citenamefont
  {Bemfica}, \citenamefont {Disconzi},\ and\ \citenamefont
  {Noronha}}]{Bemfica:2020zjp}%
  \BibitemOpen
  \bibfield  {author} {\bibinfo {author} {\bibfnamefont {Fabio~S.}\
  \bibnamefont {Bemfica}}, \bibinfo {author} {\bibfnamefont {Marcelo~M.}\
  \bibnamefont {Disconzi}}, \ and\ \bibinfo {author} {\bibfnamefont {Jorge}\
  \bibnamefont {Noronha}},\ }\bibfield  {title} {\enquote {\bibinfo {title}
  {{First-Order General-Relativistic Viscous Fluid Dynamics}},}\ }\href
  {\doibase 10.1103/PhysRevX.12.021044} {\bibfield  {journal} {\bibinfo
  {journal} {Phys. Rev. X}\ }\textbf {\bibinfo {volume} {12}},\ \bibinfo
  {pages} {021044} (\bibinfo {year} {2022})},\ \Eprint
  {http://arxiv.org/abs/2009.11388} {arXiv:2009.11388 [gr-qc]} \BibitemShut
  {NoStop}%
\bibitem [{\citenamefont {Kovtun}(2019)}]{Kovtun:2019hdm}%
  \BibitemOpen
  \bibfield  {author} {\bibinfo {author} {\bibfnamefont {Pavel}\ \bibnamefont
  {Kovtun}},\ }\bibfield  {title} {\enquote {\bibinfo {title} {{First-order
  relativistic hydrodynamics is stable}},}\ }\href {\doibase
  10.1007/JHEP10(2019)034} {\bibfield  {journal} {\bibinfo  {journal} {JHEP}\
  }\textbf {\bibinfo {volume} {10}},\ \bibinfo {pages} {034} (\bibinfo {year}
  {2019})},\ \Eprint {http://arxiv.org/abs/1907.08191} {arXiv:1907.08191
  [hep-th]} \BibitemShut {NoStop}%
\bibitem [{\citenamefont {Hoult}\ and\ \citenamefont
  {Kovtun}(2020)}]{Hoult:2020eho}%
  \BibitemOpen
  \bibfield  {author} {\bibinfo {author} {\bibfnamefont {Raphael~E.}\
  \bibnamefont {Hoult}}\ and\ \bibinfo {author} {\bibfnamefont {Pavel}\
  \bibnamefont {Kovtun}},\ }\bibfield  {title} {\enquote {\bibinfo {title}
  {{Stable and causal relativistic Navier-Stokes equations}},}\ }\href
  {\doibase 10.1007/JHEP06(2020)067} {\bibfield  {journal} {\bibinfo  {journal}
  {JHEP}\ }\textbf {\bibinfo {volume} {06}},\ \bibinfo {pages} {067} (\bibinfo
  {year} {2020})},\ \Eprint {http://arxiv.org/abs/2004.04102} {arXiv:2004.04102
  [hep-th]} \BibitemShut {NoStop}%
\bibitem [{\citenamefont {Bekenstein}(1973)}]{Bekenstein:1973mi}%
  \BibitemOpen
  \bibfield  {author} {\bibinfo {author} {\bibfnamefont {J.~D.}\ \bibnamefont
  {Bekenstein}},\ }\bibfield  {title} {\enquote {\bibinfo {title} {{Extraction
  of energy and charge from a black hole}},}\ }\href {\doibase
  10.1103/PhysRevD.7.949} {\bibfield  {journal} {\bibinfo  {journal} {Phys.
  Rev. D}\ }\textbf {\bibinfo {volume} {7}},\ \bibinfo {pages} {949--953}
  (\bibinfo {year} {1973})}\BibitemShut {NoStop}%
\bibitem [{\citenamefont {Maggio}\ \emph {et~al.}(2017)\citenamefont {Maggio},
  \citenamefont {Pani},\ and\ \citenamefont {Ferrari}}]{Maggio:2017ivp}%
  \BibitemOpen
  \bibfield  {author} {\bibinfo {author} {\bibfnamefont {Elisa}\ \bibnamefont
  {Maggio}}, \bibinfo {author} {\bibfnamefont {Paolo}\ \bibnamefont {Pani}}, \
  and\ \bibinfo {author} {\bibfnamefont {Valeria}\ \bibnamefont {Ferrari}},\
  }\bibfield  {title} {\enquote {\bibinfo {title} {{Exotic Compact Objects and
  How to Quench their Ergoregion Instability}},}\ }\href {\doibase
  10.1103/PhysRevD.96.104047} {\bibfield  {journal} {\bibinfo  {journal} {Phys.
  Rev. D}\ }\textbf {\bibinfo {volume} {96}},\ \bibinfo {pages} {104047}
  (\bibinfo {year} {2017})},\ \Eprint {http://arxiv.org/abs/1703.03696}
  {arXiv:1703.03696 [gr-qc]} \BibitemShut {NoStop}%
\bibitem [{\citenamefont {Maggio}\ \emph {et~al.}(2021)\citenamefont {Maggio},
  \citenamefont {van~de Meent},\ and\ \citenamefont {Pani}}]{Maggio:2021uge}%
  \BibitemOpen
  \bibfield  {author} {\bibinfo {author} {\bibfnamefont {Elisa}\ \bibnamefont
  {Maggio}}, \bibinfo {author} {\bibfnamefont {Maarten}\ \bibnamefont {van~de
  Meent}}, \ and\ \bibinfo {author} {\bibfnamefont {Paolo}\ \bibnamefont
  {Pani}},\ }\bibfield  {title} {\enquote {\bibinfo {title} {{Extreme
  mass-ratio inspirals around a spinning horizonless compact object}},}\ }\href
  {\doibase 10.1103/PhysRevD.104.104026} {\bibfield  {journal} {\bibinfo
  {journal} {Phys. Rev. D}\ }\textbf {\bibinfo {volume} {104}},\ \bibinfo
  {pages} {104026} (\bibinfo {year} {2021})},\ \Eprint
  {http://arxiv.org/abs/2106.07195} {arXiv:2106.07195 [gr-qc]} \BibitemShut
  {NoStop}%
\bibitem [{\citenamefont {Mathur}(2005)}]{Mathur:2005zp}%
  \BibitemOpen
  \bibfield  {author} {\bibinfo {author} {\bibfnamefont {Samir~D.}\
  \bibnamefont {Mathur}},\ }\bibfield  {title} {\enquote {\bibinfo {title}
  {{The Fuzzball proposal for black holes: An Elementary review}},}\ }\href
  {\doibase 10.1002/prop.200410203} {\bibfield  {journal} {\bibinfo  {journal}
  {Fortsch. Phys.}\ }\textbf {\bibinfo {volume} {53}},\ \bibinfo {pages}
  {793--827} (\bibinfo {year} {2005})},\ \Eprint
  {http://arxiv.org/abs/hep-th/0502050} {arXiv:hep-th/0502050} \BibitemShut
  {NoStop}%
\bibitem [{\citenamefont {Bah}\ and\ \citenamefont
  {Heidmann}(2021)}]{Bah:2020ogh}%
  \BibitemOpen
  \bibfield  {author} {\bibinfo {author} {\bibfnamefont {Ibrahima}\
  \bibnamefont {Bah}}\ and\ \bibinfo {author} {\bibfnamefont {Pierre}\
  \bibnamefont {Heidmann}},\ }\bibfield  {title} {\enquote {\bibinfo {title}
  {{Topological Stars and Black Holes}},}\ }\href {\doibase
  10.1103/PhysRevLett.126.151101} {\bibfield  {journal} {\bibinfo  {journal}
  {Phys. Rev. Lett.}\ }\textbf {\bibinfo {volume} {126}},\ \bibinfo {pages}
  {151101} (\bibinfo {year} {2021})},\ \Eprint
  {http://arxiv.org/abs/2011.08851} {arXiv:2011.08851 [hep-th]} \BibitemShut
  {NoStop}%
\bibitem [{\citenamefont {Kovtun}\ \emph {et~al.}(2003)\citenamefont {Kovtun},
  \citenamefont {Son},\ and\ \citenamefont {Starinets}}]{Kovtun:2003wp}%
  \BibitemOpen
  \bibfield  {author} {\bibinfo {author} {\bibfnamefont {Pavel}\ \bibnamefont
  {Kovtun}}, \bibinfo {author} {\bibfnamefont {Dam~T.}\ \bibnamefont {Son}}, \
  and\ \bibinfo {author} {\bibfnamefont {Andrei~O.}\ \bibnamefont
  {Starinets}},\ }\bibfield  {title} {\enquote {\bibinfo {title} {{Holography
  and hydrodynamics: Diffusion on stretched horizons}},}\ }\href {\doibase
  10.1088/1126-6708/2003/10/064} {\bibfield  {journal} {\bibinfo  {journal}
  {JHEP}\ }\textbf {\bibinfo {volume} {10}},\ \bibinfo {pages} {064} (\bibinfo
  {year} {2003})},\ \Eprint {http://arxiv.org/abs/hep-th/0309213}
  {arXiv:hep-th/0309213} \BibitemShut {NoStop}%
\bibitem [{\citenamefont {Kovtun}\ \emph {et~al.}(2005)\citenamefont {Kovtun},
  \citenamefont {Son},\ and\ \citenamefont {Starinets}}]{Kovtun:2004de}%
  \BibitemOpen
  \bibfield  {author} {\bibinfo {author} {\bibfnamefont {P.}~\bibnamefont
  {Kovtun}}, \bibinfo {author} {\bibfnamefont {Dan~T.}\ \bibnamefont {Son}}, \
  and\ \bibinfo {author} {\bibfnamefont {Andrei~O.}\ \bibnamefont
  {Starinets}},\ }\bibfield  {title} {\enquote {\bibinfo {title} {{Viscosity in
  strongly interacting quantum field theories from black hole physics}},}\
  }\href {\doibase 10.1103/PhysRevLett.94.111601} {\bibfield  {journal}
  {\bibinfo  {journal} {Phys. Rev. Lett.}\ }\textbf {\bibinfo {volume} {94}},\
  \bibinfo {pages} {111601} (\bibinfo {year} {2005})},\ \Eprint
  {http://arxiv.org/abs/hep-th/0405231} {arXiv:hep-th/0405231} \BibitemShut
  {NoStop}%
\bibitem [{\citenamefont {Hartle}(1967)}]{Hartle:1967he}%
  \BibitemOpen
  \bibfield  {author} {\bibinfo {author} {\bibfnamefont {James~B.}\
  \bibnamefont {Hartle}},\ }\bibfield  {title} {\enquote {\bibinfo {title}
  {{Slowly rotating relativistic stars. 1. Equations of structure}},}\ }\href
  {\doibase 10.1086/149400} {\bibfield  {journal} {\bibinfo  {journal}
  {Astrophys. J.}\ }\textbf {\bibinfo {volume} {150}},\ \bibinfo {pages}
  {1005--1029} (\bibinfo {year} {1967})}\BibitemShut {NoStop}%
\bibitem [{\citenamefont {Hartle}\ and\ \citenamefont
  {Thorne}(1968)}]{Hartle:1968si}%
  \BibitemOpen
  \bibfield  {author} {\bibinfo {author} {\bibfnamefont {James~B.}\
  \bibnamefont {Hartle}}\ and\ \bibinfo {author} {\bibfnamefont {Kip~S.}\
  \bibnamefont {Thorne}},\ }\bibfield  {title} {\enquote {\bibinfo {title}
  {{Slowly Rotating Relativistic Stars. II. Models for Neutron Stars and
  Supermassive Stars}},}\ }\href {\doibase 10.1086/149707} {\bibfield
  {journal} {\bibinfo  {journal} {Astrophys. J.}\ }\textbf {\bibinfo {volume}
  {153}},\ \bibinfo {pages} {807} (\bibinfo {year} {1968})}\BibitemShut
  {NoStop}%
\bibitem [{\citenamefont {Chandrasekhar}\ and\ \citenamefont
  {Ferrari}(1991)}]{Chandrasekhar:1991rmo}%
  \BibitemOpen
  \bibfield  {author} {\bibinfo {author} {\bibfnamefont {Subrahmanyan}\
  \bibnamefont {Chandrasekhar}}\ and\ \bibinfo {author} {\bibfnamefont
  {Valeria}\ \bibnamefont {Ferrari}},\ }\bibfield  {title} {\enquote {\bibinfo
  {title} {{On the non-radial oscillations of slowly rotating stars induced by
  the Lense-Thirring effect}},}\ }\href {\doibase 10.1098/rspa.1991.0056}
  {\bibfield  {journal} {\bibinfo  {journal} {Proc. Roy. Soc. Lond. A}\
  }\textbf {\bibinfo {volume} {433}},\ \bibinfo {pages} {423--440} (\bibinfo
  {year} {1991})}\BibitemShut {NoStop}%
\bibitem [{\citenamefont {Kojima}(1992)}]{Kojima:1992ie}%
  \BibitemOpen
  \bibfield  {author} {\bibinfo {author} {\bibfnamefont {Y.}~\bibnamefont
  {Kojima}},\ }\bibfield  {title} {\enquote {\bibinfo {title} {{Equations
  governing the nonradial oscillations of a slowly rotating relativistic
  star}},}\ }\href {\doibase 10.1103/PhysRevD.46.4289} {\bibfield  {journal}
  {\bibinfo  {journal} {Phys. Rev. D}\ }\textbf {\bibinfo {volume} {46}},\
  \bibinfo {pages} {4289--4303} (\bibinfo {year} {1992})}\BibitemShut {NoStop}%
\bibitem [{\citenamefont {Pani}(2013)}]{Pani:2013pma}%
  \BibitemOpen
  \bibfield  {author} {\bibinfo {author} {\bibfnamefont {Paolo}\ \bibnamefont
  {Pani}},\ }\bibfield  {title} {\enquote {\bibinfo {title} {{Advanced Methods
  in Black-Hole Perturbation Theory}},}\ }\href {\doibase
  10.1142/S0217751X13400186} {\bibfield  {journal} {\bibinfo  {journal} {Int.
  J. Mod. Phys. A}\ }\textbf {\bibinfo {volume} {28}},\ \bibinfo {pages}
  {1340018} (\bibinfo {year} {2013})},\ \Eprint
  {http://arxiv.org/abs/1305.6759} {arXiv:1305.6759 [gr-qc]} \BibitemShut
  {NoStop}%
\bibitem [{\citenamefont {Israel}(1966)}]{Israel:1966rt}%
  \BibitemOpen
  \bibfield  {author} {\bibinfo {author} {\bibfnamefont {W.}~\bibnamefont
  {Israel}},\ }\bibfield  {title} {\enquote {\bibinfo {title} {{Singular
  hypersurfaces and thin shells in general relativity}},}\ }\href {\doibase
  10.1007/BF02710419} {\bibfield  {journal} {\bibinfo  {journal} {Nuovo Cim.
  B}\ }\textbf {\bibinfo {volume} {44S10}},\ \bibinfo {pages} {1} (\bibinfo
  {year} {1966})},\ \bibinfo {note} {[Erratum: Nuovo Cim.B 48, 463
  (1967)]}\BibitemShut {NoStop}%
\bibitem [{\citenamefont {Rackauckas}\ and\ \citenamefont
  {Nie}(2017)}]{rackauckas2017differentialequations}%
  \BibitemOpen
  \bibfield  {author} {\bibinfo {author} {\bibfnamefont {Christopher}\
  \bibnamefont {Rackauckas}}\ and\ \bibinfo {author} {\bibfnamefont {Qing}\
  \bibnamefont {Nie}},\ }\bibfield  {title} {\enquote {\bibinfo {title}
  {Differential{E}quations.jl--a performant and feature-rich ecosystem for
  solving differential equations in {J}ulia},}\ }\href@noop {} {\bibfield
  {journal} {\bibinfo  {journal} {Journal of Open Research Software}\ }\textbf
  {\bibinfo {volume} {5}} (\bibinfo {year} {2017})}\BibitemShut {NoStop}%
\bibitem [{\citenamefont {Carlsson}\ \emph {et~al.}(2019)\citenamefont
  {Carlsson}, \citenamefont {Mogensen}, \citenamefont {Riseth}, \citenamefont
  {Schauer}, \citenamefont {Johansen}, \citenamefont {Udell}, \citenamefont
  {Piibeleht}, \citenamefont {Widmann}, \citenamefont {Revels}, \citenamefont
  {Noack}, \citenamefont {Kelman},\ and\ \citenamefont {Lubin}}]{nlsolve2019}%
  \BibitemOpen
  \bibfield  {author} {\bibinfo {author} {\bibfnamefont {Kristoffer}\
  \bibnamefont {Carlsson}}, \bibinfo {author} {\bibfnamefont {Patrick~Kofod}\
  \bibnamefont {Mogensen}}, \bibinfo {author} {\bibfnamefont {Asbjørn~Nilsen}\
  \bibnamefont {Riseth}}, \bibinfo {author} {\bibfnamefont {Moritz}\
  \bibnamefont {Schauer}}, \bibinfo {author} {\bibfnamefont {Jeppe}\
  \bibnamefont {Johansen}}, \bibinfo {author} {\bibfnamefont {Madeleine}\
  \bibnamefont {Udell}}, \bibinfo {author} {\bibfnamefont {Morten}\
  \bibnamefont {Piibeleht}}, \bibinfo {author} {\bibfnamefont {David}\
  \bibnamefont {Widmann}}, \bibinfo {author} {\bibfnamefont {Jarrett}\
  \bibnamefont {Revels}}, \bibinfo {author} {\bibfnamefont {Andreas}\
  \bibnamefont {Noack}}, \bibinfo {author} {\bibfnamefont {Tony}\ \bibnamefont
  {Kelman}}, \ and\ \bibinfo {author} {\bibfnamefont {Miles}\ \bibnamefont
  {Lubin}},\ }\href {\doibase 10.5281/zenodo.2682214} {\enquote {\bibinfo
  {title} {Julianlsolvers/nlsolve.jl: v4.0.0},}\ } (\bibinfo {year}
  {2019})\BibitemShut {NoStop}%
\bibitem [{web()}]{web:CoG}%
  \BibitemOpen
  \href@noop {} {}\bibinfo {note} {{Check the Center of Gravity webpage for
  publicly available material: \\ \url{https://the-center-of-gravity.com/}
  }}\BibitemShut {NoStop}%
\bibitem [{\citenamefont {Press}(1979)}]{Press:1979rd}%
  \BibitemOpen
  \bibfield  {author} {\bibinfo {author} {\bibfnamefont {W.~H.}\ \bibnamefont
  {Press}},\ }\bibfield  {title} {\enquote {\bibinfo {title} {{ON GRAVITATIONAL
  CONDUCTORS, WAVEGUIDES, AND CIRCUITS}},}\ }\href {\doibase
  10.1007/BF00756582} {\bibfield  {journal} {\bibinfo  {journal} {Gen. Rel.
  Grav.}\ }\textbf {\bibinfo {volume} {11}},\ \bibinfo {pages} {105--109}
  (\bibinfo {year} {1979})}\BibitemShut {NoStop}%
\bibitem [{\citenamefont {Saketh}\ and\ \citenamefont
  {Maggio}(2024)}]{Saketh:2024ojw}%
  \BibitemOpen
  \bibfield  {author} {\bibinfo {author} {\bibfnamefont {M.~V.~S.}\
  \bibnamefont {Saketh}}\ and\ \bibinfo {author} {\bibfnamefont {Elisa}\
  \bibnamefont {Maggio}},\ }\bibfield  {title} {\enquote {\bibinfo {title}
  {{Quasinormal modes of slowly-spinning horizonless compact objects}},}\
  }\href {\doibase 10.1103/PhysRevD.110.084038} {\bibfield  {journal} {\bibinfo
   {journal} {Phys. Rev. D}\ }\textbf {\bibinfo {volume} {110}},\ \bibinfo
  {pages} {084038} (\bibinfo {year} {2024})},\ \Eprint
  {http://arxiv.org/abs/2406.10070} {arXiv:2406.10070 [gr-qc]} \BibitemShut
  {NoStop}%
\bibitem [{\citenamefont {Cardoso}\ \emph {et~al.}(2017)\citenamefont
  {Cardoso}, \citenamefont {Pani},\ and\ \citenamefont {Yu}}]{Cardoso:2017kgn}%
  \BibitemOpen
  \bibfield  {author} {\bibinfo {author} {\bibfnamefont {Vitor}\ \bibnamefont
  {Cardoso}}, \bibinfo {author} {\bibfnamefont {Paolo}\ \bibnamefont {Pani}}, \
  and\ \bibinfo {author} {\bibfnamefont {Tien-Tien}\ \bibnamefont {Yu}},\
  }\bibfield  {title} {\enquote {\bibinfo {title} {{Superradiance in rotating
  stars and pulsar-timing constraints on dark photons}},}\ }\href {\doibase
  10.1103/PhysRevD.95.124056} {\bibfield  {journal} {\bibinfo  {journal} {Phys.
  Rev. D}\ }\textbf {\bibinfo {volume} {95}},\ \bibinfo {pages} {124056}
  (\bibinfo {year} {2017})},\ \Eprint {http://arxiv.org/abs/1704.06151}
  {arXiv:1704.06151 [gr-qc]} \BibitemShut {NoStop}%
\bibitem [{\citenamefont {Cardoso}\ \emph {et~al.}(2015)\citenamefont
  {Cardoso}, \citenamefont {Brito},\ and\ \citenamefont
  {Rosa}}]{Cardoso:2015zqa}%
  \BibitemOpen
  \bibfield  {author} {\bibinfo {author} {\bibfnamefont {Vitor}\ \bibnamefont
  {Cardoso}}, \bibinfo {author} {\bibfnamefont {Richard}\ \bibnamefont
  {Brito}}, \ and\ \bibinfo {author} {\bibfnamefont {Joao~L.}\ \bibnamefont
  {Rosa}},\ }\bibfield  {title} {\enquote {\bibinfo {title} {{Superradiance in
  stars}},}\ }\href {\doibase 10.1103/PhysRevD.91.124026} {\bibfield  {journal}
  {\bibinfo  {journal} {Phys. Rev. D}\ }\textbf {\bibinfo {volume} {91}},\
  \bibinfo {pages} {124026} (\bibinfo {year} {2015})},\ \Eprint
  {http://arxiv.org/abs/1505.05509} {arXiv:1505.05509 [gr-qc]} \BibitemShut
  {NoStop}%
\bibitem [{\citenamefont {Pani}\ \emph {et~al.}(2012)\citenamefont {Pani},
  \citenamefont {Cardoso}, \citenamefont {Gualtieri}, \citenamefont {Berti},\
  and\ \citenamefont {Ishibashi}}]{Pani:2012bp}%
  \BibitemOpen
  \bibfield  {author} {\bibinfo {author} {\bibfnamefont {Paolo}\ \bibnamefont
  {Pani}}, \bibinfo {author} {\bibfnamefont {Vitor}\ \bibnamefont {Cardoso}},
  \bibinfo {author} {\bibfnamefont {Leonardo}\ \bibnamefont {Gualtieri}},
  \bibinfo {author} {\bibfnamefont {Emanuele}\ \bibnamefont {Berti}}, \ and\
  \bibinfo {author} {\bibfnamefont {Akihiro}\ \bibnamefont {Ishibashi}},\
  }\bibfield  {title} {\enquote {\bibinfo {title} {{Perturbations of slowly
  rotating black holes: massive vector fields in the Kerr metric}},}\ }\href
  {\doibase 10.1103/PhysRevD.86.104017} {\bibfield  {journal} {\bibinfo
  {journal} {Phys. Rev. D}\ }\textbf {\bibinfo {volume} {86}},\ \bibinfo
  {pages} {104017} (\bibinfo {year} {2012})},\ \Eprint
  {http://arxiv.org/abs/1209.0773} {arXiv:1209.0773 [gr-qc]} \BibitemShut
  {NoStop}%
\bibitem [{\citenamefont {Dias}\ \emph {et~al.}(2008)\citenamefont {Dias},
  \citenamefont {Emparan},\ and\ \citenamefont {Maccarrone}}]{Dias:2007nj}%
  \BibitemOpen
  \bibfield  {author} {\bibinfo {author} {\bibfnamefont {Oscar J.~C.}\
  \bibnamefont {Dias}}, \bibinfo {author} {\bibfnamefont {Roberto}\
  \bibnamefont {Emparan}}, \ and\ \bibinfo {author} {\bibfnamefont
  {Alessandro}\ \bibnamefont {Maccarrone}},\ }\bibfield  {title} {\enquote
  {\bibinfo {title} {{Microscopic theory of black hole superradiance}},}\
  }\href {\doibase 10.1103/PhysRevD.77.064018} {\bibfield  {journal} {\bibinfo
  {journal} {Phys. Rev. D}\ }\textbf {\bibinfo {volume} {77}},\ \bibinfo
  {pages} {064018} (\bibinfo {year} {2008})},\ \Eprint
  {http://arxiv.org/abs/0712.0791} {arXiv:0712.0791 [hep-th]} \BibitemShut
  {NoStop}%
\bibitem [{\citenamefont {Cunha}\ \emph {et~al.}(2017)\citenamefont {Cunha},
  \citenamefont {Berti},\ and\ \citenamefont {Herdeiro}}]{Cunha:2017qtt}%
  \BibitemOpen
  \bibfield  {author} {\bibinfo {author} {\bibfnamefont {Pedro V.~P.}\
  \bibnamefont {Cunha}}, \bibinfo {author} {\bibfnamefont {Emanuele}\
  \bibnamefont {Berti}}, \ and\ \bibinfo {author} {\bibfnamefont {Carlos
  A.~R.}\ \bibnamefont {Herdeiro}},\ }\bibfield  {title} {\enquote {\bibinfo
  {title} {{Light-Ring Stability for Ultracompact Objects}},}\ }\href {\doibase
  10.1103/PhysRevLett.119.251102} {\bibfield  {journal} {\bibinfo  {journal}
  {Phys. Rev. Lett.}\ }\textbf {\bibinfo {volume} {119}},\ \bibinfo {pages}
  {251102} (\bibinfo {year} {2017})},\ \Eprint
  {http://arxiv.org/abs/1708.04211} {arXiv:1708.04211 [gr-qc]} \BibitemShut
  {NoStop}%
\bibitem [{\citenamefont {Keir}(2016)}]{Keir:2014oka}%
  \BibitemOpen
  \bibfield  {author} {\bibinfo {author} {\bibfnamefont {Joe}\ \bibnamefont
  {Keir}},\ }\bibfield  {title} {\enquote {\bibinfo {title} {{Slowly decaying
  waves on spherically symmetric spacetimes and ultracompact neutron stars}},}\
  }\href {\doibase 10.1088/0264-9381/33/13/135009} {\bibfield  {journal}
  {\bibinfo  {journal} {Class. Quant. Grav.}\ }\textbf {\bibinfo {volume}
  {33}},\ \bibinfo {pages} {135009} (\bibinfo {year} {2016})},\ \Eprint
  {http://arxiv.org/abs/1404.7036} {arXiv:1404.7036 [gr-qc]} \BibitemShut
  {NoStop}%
\bibitem [{\citenamefont {Cardoso}\ \emph {et~al.}(2014)\citenamefont
  {Cardoso}, \citenamefont {Crispino}, \citenamefont {Macedo}, \citenamefont
  {Okawa},\ and\ \citenamefont {Pani}}]{Cardoso:2014sna}%
  \BibitemOpen
  \bibfield  {author} {\bibinfo {author} {\bibfnamefont {Vitor}\ \bibnamefont
  {Cardoso}}, \bibinfo {author} {\bibfnamefont {Lu\'\i{}s C.~B.}\ \bibnamefont
  {Crispino}}, \bibinfo {author} {\bibfnamefont {Caio F.~B.}\ \bibnamefont
  {Macedo}}, \bibinfo {author} {\bibfnamefont {Hirotada}\ \bibnamefont
  {Okawa}}, \ and\ \bibinfo {author} {\bibfnamefont {Paolo}\ \bibnamefont
  {Pani}},\ }\bibfield  {title} {\enquote {\bibinfo {title} {{Light rings as
  observational evidence for event horizons: long-lived modes, ergoregions and
  nonlinear instabilities of ultracompact objects}},}\ }\href {\doibase
  10.1103/PhysRevD.90.044069} {\bibfield  {journal} {\bibinfo  {journal} {Phys.
  Rev. D}\ }\textbf {\bibinfo {volume} {90}},\ \bibinfo {pages} {044069}
  (\bibinfo {year} {2014})},\ \Eprint {http://arxiv.org/abs/1406.5510}
  {arXiv:1406.5510 [gr-qc]} \BibitemShut {NoStop}%
\bibitem [{\citenamefont {Zhong}\ \emph {et~al.}(2023)\citenamefont {Zhong},
  \citenamefont {Cardoso},\ and\ \citenamefont {Maggio}}]{Zhong:2022jke}%
  \BibitemOpen
  \bibfield  {author} {\bibinfo {author} {\bibfnamefont {Zhen}\ \bibnamefont
  {Zhong}}, \bibinfo {author} {\bibfnamefont {Vitor}\ \bibnamefont {Cardoso}},
  \ and\ \bibinfo {author} {\bibfnamefont {Elisa}\ \bibnamefont {Maggio}},\
  }\bibfield  {title} {\enquote {\bibinfo {title} {{Instability of ultracompact
  horizonless spacetimes}},}\ }\href {\doibase 10.1103/PhysRevD.107.044035}
  {\bibfield  {journal} {\bibinfo  {journal} {Phys. Rev. D}\ }\textbf {\bibinfo
  {volume} {107}},\ \bibinfo {pages} {044035} (\bibinfo {year} {2023})},\
  \Eprint {http://arxiv.org/abs/2211.16526} {arXiv:2211.16526 [gr-qc]}
  \BibitemShut {NoStop}%
\bibitem [{\citenamefont {Benomio}\ \emph {et~al.}(2024)\citenamefont
  {Benomio}, \citenamefont {C\'ardenas-Avenda\~no}, \citenamefont {Pretorius},\
  and\ \citenamefont {Sullivan}}]{Benomio:2024lev}%
  \BibitemOpen
  \bibfield  {author} {\bibinfo {author} {\bibfnamefont {Gabriele}\
  \bibnamefont {Benomio}}, \bibinfo {author} {\bibfnamefont {Alejandro}\
  \bibnamefont {C\'ardenas-Avenda\~no}}, \bibinfo {author} {\bibfnamefont
  {Frans}\ \bibnamefont {Pretorius}}, \ and\ \bibinfo {author} {\bibfnamefont
  {Andrew}\ \bibnamefont {Sullivan}},\ }\bibfield  {title} {\enquote {\bibinfo
  {title} {{On turbulence for spacetimes with stable trapping}},}\ }\href@noop
  {} {\  (\bibinfo {year} {2024})},\ \Eprint {http://arxiv.org/abs/2411.17445}
  {arXiv:2411.17445 [gr-qc]} \BibitemShut {NoStop}%
\bibitem [{\citenamefont {Redondo-Yuste}\ and\ \citenamefont
  {C\'ardenas-Avenda\~no}(2025)}]{Redondo-Yuste:2025hlv}%
  \BibitemOpen
  \bibfield  {author} {\bibinfo {author} {\bibfnamefont {Jaime}\ \bibnamefont
  {Redondo-Yuste}}\ and\ \bibinfo {author} {\bibfnamefont {Alejandro}\
  \bibnamefont {C\'ardenas-Avenda\~no}},\ }\bibfield  {title} {\enquote
  {\bibinfo {title} {{Perturbative and non-linear analyses of gravitational
  turbulence in spacetimes with stable light rings}},}\ }\href@noop {} {\
  (\bibinfo {year} {2025})},\ \Eprint {http://arxiv.org/abs/2502.18643}
  {arXiv:2502.18643 [gr-qc]} \BibitemShut {NoStop}%
\bibitem [{\citenamefont {Marks}\ \emph {et~al.}(2025)\citenamefont {Marks},
  \citenamefont {Staelens}, \citenamefont {Evstafyeva},\ and\ \citenamefont
  {Sperhake}}]{Marks:2025jpt}%
  \BibitemOpen
  \bibfield  {author} {\bibinfo {author} {\bibfnamefont {Gareth~Arturo}\
  \bibnamefont {Marks}}, \bibinfo {author} {\bibfnamefont {Seppe~J.}\
  \bibnamefont {Staelens}}, \bibinfo {author} {\bibfnamefont {Tamara}\
  \bibnamefont {Evstafyeva}}, \ and\ \bibinfo {author} {\bibfnamefont {Ulrich}\
  \bibnamefont {Sperhake}},\ }\bibfield  {title} {\enquote {\bibinfo {title}
  {{Long-term stable nonlinear evolutions of ultracompact black-hole
  mimickers}},}\ }\href@noop {} {\  (\bibinfo {year} {2025})},\ \Eprint
  {http://arxiv.org/abs/2504.17775} {arXiv:2504.17775 [gr-qc]} \BibitemShut
  {NoStop}%
\bibitem [{\citenamefont {Teukolsky}\ and\ \citenamefont
  {Press}(1974)}]{Teukolsky:1974yv}%
  \BibitemOpen
  \bibfield  {author} {\bibinfo {author} {\bibfnamefont {S.~A.}\ \bibnamefont
  {Teukolsky}}\ and\ \bibinfo {author} {\bibfnamefont {W.~H.}\ \bibnamefont
  {Press}},\ }\bibfield  {title} {\enquote {\bibinfo {title} {{Perturbations of
  a rotating black hole. III - Interaction of the hole with gravitational and
  electromagnet ic radiation}},}\ }\href {\doibase 10.1086/153180} {\bibfield
  {journal} {\bibinfo  {journal} {Astrophys. J.}\ }\textbf {\bibinfo {volume}
  {193}},\ \bibinfo {pages} {443--461} (\bibinfo {year} {1974})}\BibitemShut
  {NoStop}%
\bibitem [{\citenamefont {Berti}\ \emph {et~al.}(2004)\citenamefont {Berti},
  \citenamefont {Cardoso},\ and\ \citenamefont {Lemos}}]{Berti:2004ju}%
  \BibitemOpen
  \bibfield  {author} {\bibinfo {author} {\bibfnamefont {Emanuele}\
  \bibnamefont {Berti}}, \bibinfo {author} {\bibfnamefont {Vitor}\ \bibnamefont
  {Cardoso}}, \ and\ \bibinfo {author} {\bibfnamefont {Jose P.~S.}\
  \bibnamefont {Lemos}},\ }\bibfield  {title} {\enquote {\bibinfo {title}
  {{Quasinormal modes and classical wave propagation in analogue black
  holes}},}\ }\href {\doibase 10.1103/PhysRevD.70.124006} {\bibfield  {journal}
  {\bibinfo  {journal} {Phys. Rev. D}\ }\textbf {\bibinfo {volume} {70}},\
  \bibinfo {pages} {124006} (\bibinfo {year} {2004})},\ \Eprint
  {http://arxiv.org/abs/gr-qc/0408099} {arXiv:gr-qc/0408099} \BibitemShut
  {NoStop}%
\bibitem [{\citenamefont {Cardoso}\ \emph {et~al.}(2016)\citenamefont
  {Cardoso}, \citenamefont {Coutant}, \citenamefont {Richartz},\ and\
  \citenamefont {Weinfurtner}}]{Cardoso:2016zvz}%
  \BibitemOpen
  \bibfield  {author} {\bibinfo {author} {\bibfnamefont {Vitor}\ \bibnamefont
  {Cardoso}}, \bibinfo {author} {\bibfnamefont {Antonin}\ \bibnamefont
  {Coutant}}, \bibinfo {author} {\bibfnamefont {Mauricio}\ \bibnamefont
  {Richartz}}, \ and\ \bibinfo {author} {\bibfnamefont {Silke}\ \bibnamefont
  {Weinfurtner}},\ }\bibfield  {title} {\enquote {\bibinfo {title} {{Detecting
  Rotational Superradiance in Fluid Laboratories}},}\ }\href {\doibase
  10.1103/PhysRevLett.117.271101} {\bibfield  {journal} {\bibinfo  {journal}
  {Phys. Rev. Lett.}\ }\textbf {\bibinfo {volume} {117}},\ \bibinfo {pages}
  {271101} (\bibinfo {year} {2016})},\ \Eprint
  {http://arxiv.org/abs/1607.01378} {arXiv:1607.01378 [gr-qc]} \BibitemShut
  {NoStop}%
\bibitem [{\citenamefont {Torres}\ \emph {et~al.}(2017)\citenamefont {Torres},
  \citenamefont {Patrick}, \citenamefont {Coutant}, \citenamefont {Richartz},
  \citenamefont {Tedford},\ and\ \citenamefont {Weinfurtner}}]{Torres:2016iee}%
  \BibitemOpen
  \bibfield  {author} {\bibinfo {author} {\bibfnamefont {Theo}\ \bibnamefont
  {Torres}}, \bibinfo {author} {\bibfnamefont {Sam}\ \bibnamefont {Patrick}},
  \bibinfo {author} {\bibfnamefont {Antonin}\ \bibnamefont {Coutant}}, \bibinfo
  {author} {\bibfnamefont {Mauricio}\ \bibnamefont {Richartz}}, \bibinfo
  {author} {\bibfnamefont {Edmund~W.}\ \bibnamefont {Tedford}}, \ and\ \bibinfo
  {author} {\bibfnamefont {Silke}\ \bibnamefont {Weinfurtner}},\ }\bibfield
  {title} {\enquote {\bibinfo {title} {{Observation of superradiance in a
  vortex flow}},}\ }\href {\doibase 10.1038/nphys4151} {\bibfield  {journal}
  {\bibinfo  {journal} {Nature Phys.}\ }\textbf {\bibinfo {volume} {13}},\
  \bibinfo {pages} {833--836} (\bibinfo {year} {2017})},\ \Eprint
  {http://arxiv.org/abs/1612.06180} {arXiv:1612.06180 [gr-qc]} \BibitemShut
  {NoStop}%
\bibitem [{\citenamefont {Bekenstein}\ and\ \citenamefont
  {Schiffer}(1998)}]{Bekenstein:1998nt}%
  \BibitemOpen
  \bibfield  {author} {\bibinfo {author} {\bibfnamefont {Jacob~D.}\
  \bibnamefont {Bekenstein}}\ and\ \bibinfo {author} {\bibfnamefont {Marcelo}\
  \bibnamefont {Schiffer}},\ }\bibfield  {title} {\enquote {\bibinfo {title}
  {{The Many faces of superradiance}},}\ }\href {\doibase
  10.1103/PhysRevD.58.064014} {\bibfield  {journal} {\bibinfo  {journal} {Phys.
  Rev. D}\ }\textbf {\bibinfo {volume} {58}},\ \bibinfo {pages} {064014}
  (\bibinfo {year} {1998})},\ \Eprint {http://arxiv.org/abs/gr-qc/9803033}
  {arXiv:gr-qc/9803033} \BibitemShut {NoStop}%
\bibitem [{\citenamefont {Cromb}\ \emph {et~al.}(2025)\citenamefont {Cromb},
  \citenamefont {Braidotti}, \citenamefont {Vinante}, \citenamefont {Faccio},\
  and\ \citenamefont {Ulbricht}}]{Cromb:2025dqu}%
  \BibitemOpen
  \bibfield  {author} {\bibinfo {author} {\bibfnamefont {Marion}\ \bibnamefont
  {Cromb}}, \bibinfo {author} {\bibfnamefont {Maria~Chiara}\ \bibnamefont
  {Braidotti}}, \bibinfo {author} {\bibfnamefont {Andrea}\ \bibnamefont
  {Vinante}}, \bibinfo {author} {\bibfnamefont {Daniele}\ \bibnamefont
  {Faccio}}, \ and\ \bibinfo {author} {\bibfnamefont {Hendrik}\ \bibnamefont
  {Ulbricht}},\ }\bibfield  {title} {\enquote {\bibinfo {title} {{Creation of a
  black hole bomb instability in an electromagnetic system}},}\ }\href@noop {}
  {\  (\bibinfo {year} {2025})},\ \Eprint {http://arxiv.org/abs/2503.24034}
  {arXiv:2503.24034 [quant-ph]} \BibitemShut {NoStop}%
\end{thebibliography}%
\clearpage

\appendix
\section{Scalar Waves}

\begin{figure}
    \centering
    \includegraphics[width=\columnwidth]{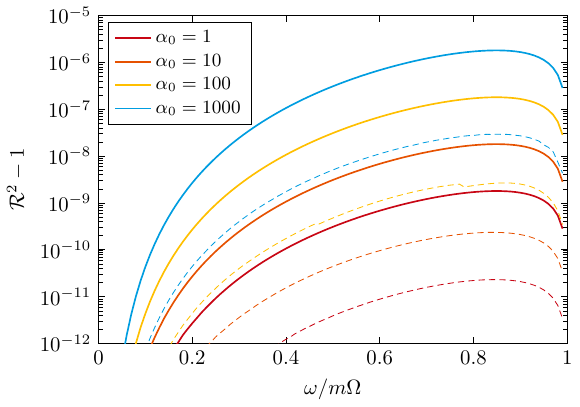}
    \caption{Reflectivity minus one $\mathcal{R}^2-1$ as a function of the dimensionless frequency $\omega/m\Omega$ for the toy model~\eqref{eq:Toy_Model}. Solid lines correspond to a constant profile of $\alpha$, which has a sharp cutoff at the surface of the star, whereas dashed lines correspond to a smooth profile of $\alpha$, which goes to zero smoothly towards the surface. The amplification factor scales proportionally with $\alpha$, as expected.}
    \label{fig:toy}
\end{figure}

Let us first study a toy model describing scalar waves propagating in the spacetime of a rotating star, with an additional term that effectively accounts for absorption. Consider the following equation for a scalar field:
\begin{equation}\label{eq:Toy_Model}
    \Box\Phi = \alpha u^\mu \nabla_\mu \Phi \, , 
\end{equation}
where $\alpha=\alpha(r)$ is a non-negative, purely radial function with units of frequency. It vanishes outside the star $\alpha(r>R_S)=0$, and $\Box$ is the wave operator in the spacetime of a slowly rotating star. Here $u^\mu$ is the fluid velocity, given in Eq.~\eqref{bg_fluid_velocity}. Expanding in spherical harmonics $r\Phi = \sum_{\ell m}\phi_{\ell m} Y_{\ell m}$, neglecting mode coupling contributions, and transforming to the frequency domain, we find
\begin{equation}
    \frac{d^2\phi}{dr_\star^2} + \Bigl(\omega^2-\mathcal{V}\Bigr)\phi = i\alpha e^{\nu/2}(\omega-m\Omega)\phi \, , 
\end{equation}
where 
\begin{equation}
    \mathcal{V} = \frac{e^\nu}{r^2}\Bigl[\ell(\ell+1)+\frac{2M}{r}+4\pi(p-\rho)\Bigr]+2m\omega\varpi  \, .
\end{equation}
For low frequencies ($\omega < m\Omega$), the damping term on the right-hand side becomes an amplification term. We integrate this equation numerically from the interior of the star, ensuring regularity near the origin, and extract the reflectivity sufficiently far away, as described in the main text. Our results are shown in Fig.~\ref{fig:toy}. We consider two models for dissipation: (i) $\alpha=\alpha_0/R_S$, a sharp cutoff at the surface of the star, and (ii) $\alpha = \alpha_0 \sqrt{\rho}$, which smoothly approaches zero at the surface. The qualitative behavior for both models is identical -- frequencies below the superradiant bound $\omega<m\Omega$ are amplified, whereas waves are absorbed by the star past this bound. Amplification increases with larger values of $\alpha$, as one may naively expect.

This simplified model also allows us to study potential instabilities in the presence of null trapping. A constant density star $\rho=\rho_c$ will have a stable light ring whenever $R_S<3M_S$ (to linear order in the rotation rate). This scalar model avoids the complicated boundary conditions present at the surface of a constant density star, which would be significantly more challenging in the gravitational case discussed in the main text. 

\begin{figure}
    \centering
    \includegraphics[width=\columnwidth]{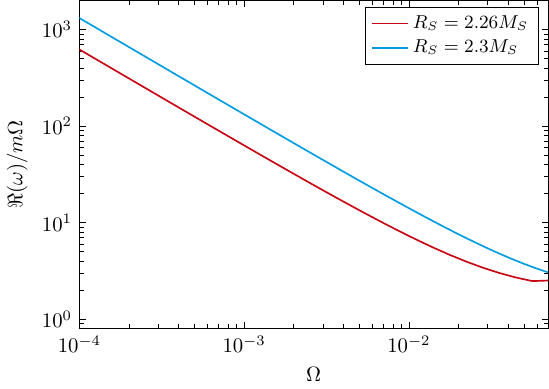}
    \caption{Real part of the frequency of the fundamental mode, rescaled by $m\Omega$, for $m=2$, as a function of the angular velocity of the star, $\Omega$, for the model of Eq.~\eqref{eq:Toy_Model}. The red (blue) lines correspond to stars with radius $R_S=2.26M_S$ ($R_S=2.3M_S$). The dimensionless frequency $\Re\omega/m\Omega > 1$ is always larger than unity, signaling that the fundamental mode is outside the superradiant amplification threshold, even at relatively large angular velocities of the star. }
    \label{fig:TOY_QNM}
\end{figure}

We adapt the previous method, and integrate the equation also from the exterior, requiring outgoing boundary conditions asymptotically far away. Then, we shoot numerically for the quasinormal mode frequencies at which the solution is (i) regular at the origin, and (ii) outgoing at large distances. By slowly changing the dissipation rate $\alpha$ and the angular velocity of the star $\Omega$ we can smoothly keep track of the evolution of the fundamental mode, shown in Fig.~\ref{fig:TOY_QNM}. As the figure shows, the real part of the frequency of the fundamental mode is always larger than the superradiant threshold, $\Re\omega >m\Omega$. In the regime where $\Omega \gg \Omega_{\rm sLR}$, the rescaled frequency $\Re\omega / (m\Omega)\to \mathscr{O}(1)$ asymptotes to a constant of order $\mathscr{O}(1)$, confirming the scaling $\omega \sim m(\Omega+\Omega_{\rm sLR})$. This behavior is independent of the compactness of the star, provided it has a stable light ring. Although not shown in the Figure, we report that the imaginary part depends only very weakly on $\Omega$. 

\section{Separating the equations}
In this Appendix, we briefly review the procedure used to separate the equations in the slowly rotating limit, as described in~\cite{Kojima:1992ie, Pani:2013pma}. Let us denote the linearized Einstein equations by $\mathcal{E}_{ab}=\delta(R_{ab}-Rg_{ab}/2-8\pi T_{ab})$. To consider the axial sector, we focus on some components, which can be expanded using spherical harmonics as follows:
\begin{equation}
    \begin{aligned}
        \mathcal{E}_{I\theta} =& \tilde{\alpha}_{\ell m}^I\cos\theta \partial_\theta Y_{\ell m}-\frac{\beta^I_{\ell m}}{\sin\theta}\partial_\phi Y_{\ell m} + \eta^I_{\ell m}\sin\theta Y_{\ell m }\\
        &+\chi^I_{\ell m}\sin\theta W_{\ell m} + \dots \, , \\
        \mathcal{E}_{I\phi} =& \beta_{\ell m}^I \partial_\theta Y_{\ell m}+\frac{\tilde{\alpha}^I_{\ell m}\cos\theta}{\sin\theta}\partial_\phi Y_{\ell m} + \chi^I_{\ell m}X_{\ell m } + \dots \, , \\
        \mathcal{E}_{\theta\phi} =& \frac{g_{\ell m}}{\sin\theta}\partial_\phi Y_{\ell m}+\frac{t_{\ell m}}{\sin\theta}W_{\ell m}+\dots \, , \\
        \mathcal{E}_- \equiv& \mathcal{E}_{\theta\theta}-\frac{\mathcal{E}_{\phi\phi}}{\sin^2\theta} = g_{\ell m}\partial_\theta Y_{\ell m}-\frac{t_{\ell m}}{\sin^2\theta}X_{\ell m}+\dots \, ,
    \end{aligned}
\end{equation}
with $I=t,r$, the dots denote even-parity terms, which we omit. We have introduced 
\begin{equation}
    \begin{aligned}
        X_{\ell m }=& 2\partial_\phi\Bigl(\partial_\theta-\cot\theta\Bigr)Y_{\ell m} \, , \\
        W_{\ell m}=&\Bigl(\partial^2_\theta-\cot\theta\partial_\theta-\frac{1}{\sin^2\theta}\partial^2_\phi\Bigr)Y_{\ell m} \, .
    \end{aligned}
\end{equation}
Projecting these equations onto odd parity spherical harmonics leads to three equations, which, after neglecting the mode coupling between even and odd parity sectors, can be written as 
\begin{equation}\label{projected_efe}
    \begin{aligned}
        \ell(\ell+1)\beta^I_{\ell m}+im\Bigl[(\ell-1)(\ell+2)\chi^I_{\ell m}+\tilde{\alpha}^I_{\ell m}+\eta^I_{\ell m}\Bigr] =& 0 \, , \\
        \ell(\ell+1)t_{\ell m}+i m g_{\ell m} =& 0 \, .
    \end{aligned}
\end{equation}
Similarly, if we let $\mathcal{C}_a = \delta\Bigl(\nabla^bT_{ab}\Bigr)$ be the linearization of the conservation of the stress-energy tensor, we can write 
\begin{equation}
    \begin{aligned}
        \mathcal{C}_{\theta} =& \hat{\alpha}_{\ell m}\cos\theta \partial_\theta Y_{\ell m}-\frac{\hat{\beta}_{\ell m}}{\sin\theta}\partial_\phi Y_{\ell m} + \hat{\eta}_{\ell m}\sin\theta Y_{\ell m }\\
        &+\hat{\chi}_{\ell m}\sin\theta W_{\ell m} + \dots \, , \\
        \mathcal{C}_{\phi} =& \hat{\beta}_{\ell m} \partial_\theta Y_{\ell m}+\frac{\hat{\alpha}_{\ell m}\cos\theta}{\sin\theta}\partial_\phi Y_{\ell m} + \hat{\chi}_{\ell m}X_{\ell m }+ \dots \, ,
    \end{aligned}
\end{equation}
leading to the equation 
\begin{equation}\label{projected_cset}
    \ell(\ell+1)\hat{\beta}_{\ell m}+im\Bigl[(\ell-1)(\ell+2)\hat{\chi}_{\ell m}+\hat{\alpha}_{\ell m}+\hat{\eta}_{\ell m}\Bigr] = 0 \, .
\end{equation}
The system of equations formed by Eqs.~\eqref{projected_efe}--\eqref{projected_cset} can now be used to derive two coupled wave equations. We first use the $t$-component of Eq.~\eqref{projected_efe} to express $h_0$ in terms of $Z$ and $
\psi$, also utilizing the $r$-component of Eq.~\eqref{projected_efe} and Eq.~\eqref{projected_cset} to eliminate $h_0'$ and $Z''$, respectively. Once we have eliminated $h_0$, we substitute back into the $r$-component of Eq.~\eqref{projected_efe}, and Eq.~\eqref{projected_cset} to obtain wave equations for $\psi$ and $Z$, respectively. The second equation of Eqs.~\eqref{projected_efe} (the angular part) is then used to verify the correctness of the derivation. These coupled wave equations take the form of Eqs.~\eqref{wave_eqs}. The coefficients of the equations are lengthy and unilluminating, so we provide them as a \texttt{Mathematica} notebook and will make them available in other formats upon request.

\section{Boundary Conditions}

The surface of the star $r=R_S$ is a spacelike boundary, where we must impose Israel junction conditions~\cite{Israel:1966rt}. These conditions imply that $[[G_{r\mu}]]=0$, where $[[X]]=X(R_S^+)-X(R_S^-)$ denotes the jump of the quantity $X$ at the surface. This is trivially satisfied for the background, as the stress-energy tensor vanishes at the surface of the star. In this case, it suffices to require that $\delta T_{r\mu}$ vanishes at the surface of the star and that $\psi,\psi'$ are continuous. At the surface of the star, $\delta T_{rA}\propto \tau_\Q\delta T_1 +\eta \delta T_2 + p \delta T_3$, where $A=\theta,\phi$, and $\delta T_{1,2,3}$ are finite at the surface. Since for the viscous parametrization and equation of state chosen, $\tau_\Q,\eta\to 0$ at the surface of the star, Israel junction conditions are trivially satisfied. 

However, the wave equation for $Z$ acquires a divergent contribution at the surface. Regular solutions are those for which this divergent piece vanishes. Analyzing the equation close to the surface, as in~\cite{Boyanov:2024jge}, results in the following boundary condition:
\begin{equation}
    A_1\frac{dZ}{dr}+A_2Z+A_3\frac{d\psi}{dr}+A_4\psi = 0 \, , 
\end{equation}
with
\begin{widetext}
\begin{equation}
    \begin{aligned}
        A_1 =&  \hat{\eta}\ell(\ell+1)M_S R_S^4 z_S^3 \, , \qquad A_3 = \ell(\ell+1)\Bigl(1-\frac{m\Omega}{\omega}\Bigr)R_S^6z_S^4 \,\\
        A_2 =& i R_S^2z_S^2\Biggl[4mJ_S+2i\ell(\ell+1)M_SR_Sz_S\hat{\eta}-R_S^3 \Bigl(\ell(\ell+1)\omega+2m\Omega - m\ell(\ell+1)\Omega\Bigr)\Biggr] \, , \\
        A_4 =& \frac{2mJ_S}{\omega}\Bigl( 6z_S^2+i M_SR_S^2z_S\ell(\ell+1)\omega\hat{\eta} \Bigr) +\ell(\ell+1)R_S^3\Bigl(1-\frac{m\Omega}{\omega}\Bigr)\Bigl(z_S^2-iM_SR_S^2z_S\omega\hat{\eta}\Bigr) \, .
    \end{aligned}
\end{equation}
\end{widetext}
\end{document}